\documentclass[a4paper,fleqn,usenatbib]{mnras}

\usepackage{newtxtext,newtxmath}

\usepackage[T1]{fontenc}
\usepackage{ae,aecompl}

\usepackage{graphicx}	
\usepackage{amsmath}	
\usepackage{amssymb}	
\usepackage{hyperref}
\usepackage{longtable}
\usepackage{caption}
\usepackage{pdflscape}
\usepackage{tabularx}
\usepackage{comment}

\title[The importance of realism]{Deep learning predictions of galaxy merger stage and the importance of observational realism}
\author[Bottrell et al.]{Connor Bottrell$^{1}$\thanks{E-mail: cbottrel@uvic.ca},
Maan H. Hani$^{1\dagger}$,
Hossen Teimoorinia$^{2,1}$,
Sara L. Ellison$^{1}$,
\newauthor
Jorge Moreno$^{3,4,5}$, 
Paul Torrey$^{6}$,
Christopher C. Hayward$^{7}$,
Mallory Thorp$^{1}$,
\newauthor
Luc Simard$^{2}$ \&
Lars Hernquist$^{4}$
\\
$^{1}$Department of Physics and Astronomy, University of Victoria, Victoria, British Columbia V8P 1A1, Canada\\
$^{2}$National Research Council of Canada, 5071 West Saanich Road, Victoria, British Columbia V9E 2E7, Canada\\
$^{3}$Department of Physics and Astronomy, Pomona College, Claremont, CA 91711, USA \\
$^{4}$Harvard-Smithsonian Center for Astrophysics, 60 Garden Street, Cambridge, MA, 02138, USA\\
$^{5}$TAPIR, Mailcode 350-17, California Institute of Technology, Pasadena, CA 91125, USA\\
$^{6}$Department of Astronomy, University of Florida, 211 Bryant Space Science Center, Gainesville, FL, 32611, USA\\
$^{7}$Center for Computational Astrophysics, Flatiron Institute, 162 Fifth Avenue, New York, NY 10010, USA\\
$^{\dagger}$Vanier Scholar
}

\date{Accepted XXX. Received YYY; in original form ZZZ}

\pubyear{2019}

\begin{document}
\label{firstpage}
\pagerange{\pageref{firstpage}--\pageref{lastpage}}
\maketitle

\begin{abstract}

Machine learning is becoming a popular tool to quantify galaxy morphologies and identify mergers. However, this technique relies on using an appropriate set of training data to be successful. By combining hydrodynamical simulations, synthetic observations and convolutional neural networks (CNNs), we quantitatively assess how realistic simulated galaxy images must be in order to reliably classify mergers. Specifically, we compare the performance of CNNs trained with two types of galaxy images, stellar maps and dust-inclusive radiatively transferred images, each with three levels of observational realism: (1) no observational effects (idealized images), (2) realistic sky and point spread function (semi-realistic images), (3) insertion into a real sky image (fully realistic images).  We find that networks trained on either idealized or semi-real images have poor performance when applied to survey-realistic images. In contrast, networks trained on fully realistic images achieve 87.1\% classification performance. Importantly, the level of realism in the training images is much more important than whether the images included radiative transfer, or simply used the stellar maps ($87.1\%$ compared to $79.6\%$ accuracy, respectively). Therefore, one can avoid the large computational and storage cost of running radiative transfer with a relatively modest compromise in classification performance. Making photometry-based networks insensitive to colour incurs a very mild penalty to performance with survey-realistic data ($86.0\%$ with $r$-only compared to $87.1\%$ with $gri$). This result demonstrates that while colour \emph{can} be exploited by colour-sensitive networks, it is not necessary to achieve high accuracy and so can be avoided if desired. We provide the public release of our statistical observational realism suite, \textsc{RealSim}, as a companion to this paper.

\end{abstract}




\section{Introduction}\label{sec:intro}

Theoretical predictions and observations alike show that mergers transform galaxies. Stellar bridges and tails observed in interacting galaxy pairs are the relics of the strong gravitational and tidal forces involved in close galaxy-galaxy encounters \citep{1972ApJ...178..623T}. But the consequences of these forces extend well beyond immediate changes to visual morphology. 

Tidal torques and shocks excited by close encounters can rapidly reduce angular momentum in the dynamically cold interstellar medium (ISM) through various channels -- all driving inflow of available cold gas towards the centres of interacting galaxies (e.g. \citealt{1989Natur.340..687H,1992ARA&A..30..705B,1996ApJ...464..641M,2010MNRAS.407.1529H,2018MNRAS.479.3952B}). There is now a strong numerical and observational framework linking this rapid and central accumulation of gas to boosts in central star formation rates (e.g. \citealt{2008AJ....135.1877E,2011MNRAS.412..591P,2013MNRAS.430.1901H,2013MNRAS.433L..59P,2015MNRAS.448.1107M,2016MNRAS.462.2418S,2019MNRAS.482L..55T}), dilution of central gas phase metallicity (e.g. \citealt{2006AJ....131.2004K,2008AJ....135.1877E,2010ApJ...710L.156R,2010ApJ...723.1255R,2010A&A...514A..57S,2011MNRAS.417..580P,2012ApJ...746..108T,2015MNRAS.448.1107M,2019MNRAS.482L..55T}) and accretion onto central black holes and triggering of active galactic nuclei (AGN, e.g. \citealt{1985AJ.....90..708K,2005Natur.433..604D,2010ApJ...716L.125K,2011MNRAS.418.2043E,2014MNRAS.441.1297S,2015MNRAS.451L..35E,2018PASJ...70S..37G,2019MNRAS.487.2491E}). Additionally, galactic outflows of gas associated with the enhancements in star-formation rates (e.g. \citealt{2005ApJ...621..227M,2005ApJS..160..115R,2009ApJ...697.2030S,2017MNRAS.465.1682H}) and AGN activity (e.g. \citealt{2005ApJ...632..751R,2013ApJ...776...27V,2016ApJ...832..142Z,2017ApJ...839..120W}) can also be triggered by mergers -- resulting in the growth and enrichment the circum-galactic medium (e.g. \citealt{2015MNRAS.449.3263J,2018MNRAS.475.1160H}). Combined with the role of mergers in the assembly of present-day galaxies (e.g. \citealt{1978MNRAS.183..341W,1984Natur.311..517B}) and transforming their morphologies and kinematics (e.g. \citealt{1977egsp.conf..401T,1983MNRAS.205.1009N,1992ApJ...400..460H,2003ApJ...597..893N,2008ApJ...679..156H,2014MNRAS.440L..66B}) these connections make mergers complex but unique laboratories for testing some of the most crucial aspects of galaxy formation physics.

One observationally measurable parameter that is particularly valuable for testing the statistical and cosmological role of mergers in galaxy evolution (and which is directly comparable to numerical predictions from semi-analytic models or cosmological hydrodynamical simulations) is the galaxy merger rate and its evolution with mass and redshift (e.g. \citealt{1993MNRAS.262..627L,2011A&A...530A..20L,2011ApJ...742..103L,2012ApJ...747...34B,2013A&A...558A.135L,2014MNRAS.445.1157C,2015MNRAS.449...49R,2018MNRAS.480.2266M}). Estimating the merger rate requires: (1) a method with which mergers can be distinguished from non-merging galaxies and (2) an estimate of the timescales on which the distinction can be made -- which is sensitive to the method used in the former \citep{2008ApJS..175..356H,2008MNRAS.391.1137L,2010MNRAS.404..575L,2010MNRAS.404..590L}. However, beyond identifying mergers, we are also particularly interested in predicting merger \emph{stage}. Hydrodynamical simulations of galaxy mergers predict significant evolution in (among others) star-formation rates, ISM content, AGN accretion rates and luminosities, and subsequent stellar and AGN feedback along the merger sequence (e.g. \citealt{2008MNRAS.384..386C,2008ApJS..175..390H,2012ApJ...746..108T,2013MNRAS.430.1901H,2015MNRAS.448.1107M,2019MNRAS.485.1320M}). Consequently, in order to test the broader and detailed elements of this framework (such as feedback and outflow prescriptions), we must be able to (i) obtain large and reasonably complete observational samples of galaxy mergers and (ii) connect observed galaxy interactions to specific stages in the merger sequence. 

Both of these tasks present significant challenges from an observational perspective. For example, while mergers and recent post-mergers can be selected visually on the basis of distinct (but often low surface-brightness) morphological features such as tidal tails, bridges, streams, shells and nearby companions \citep{2010MNRAS.401.1043D,2015ApJS..221...11K,2017MNRAS.464.4420S}, this process is subjective and sensitive to contrast, resolution and surface-brightness limits. For pair candidates, the intrinsic subjectivity of visual classification can be alleviated by obtaining relative velocities with spectroscopy. Spectroscopic pair identification is effective even at high-redshifts \citep{2007ApJ...660L..51L,2011ApJ...728..119W}, but is often incomplete due to the ``fibre-collision'' problem and sparse sampling -- which particularly affect close pair completeness (\citealt{2002ApJ...565..208P,2004ApJ...617L...9L,2008ApJ...685..235P}, but see also \citealt{2014MNRAS.444.3986R}). Fast and reproducible classifications can be made using automated quantitative morphologies \citep{2003ApJS..147....1C,2004AJ....128..163L,2016MNRAS.456.3032P,2019MNRAS.483.4140R}. These metrics are designed to exploit the excess asymmetries, disturbed morphologies or multiple nuclei of mergers and merger-remnants relative to non-merging galaxies (e.g. \citealt{2013MNRAS.429.1051C,2014MNRAS.445.1157C,2016MNRAS.461.2589P}). The main obstacle with quantitative morphologies is defining empirical thresholds which separate merger from non-merger classes. Like visual classification, these thresholds (particularly for asymmetries) are sensitive to resolution and surface brightness limits (e.g. \citealt{2014A&A...566A..97J,2019MNRAS.486..390B}) but also, critically, the ``training'' data with which these thresholds are calibrated. 

To calibrate an empirical threshold for a metric which separates merger from non-merger classes, one must have a way of evaluating its performance (e.g. completeness and/or purity). The significant limitation of calibrating on observational data is that the subjective and non-subjective biases afflicting visual classifications and the incompleteness of spectroscopic samples become embedded in the calibration. In other words, observationally, one does not have access to the ground truth. This limitation can be overcome using synthetic images from hydrodynamical merger simulations as the basis for the calibration step -- which simultaneously solves the problems above \citep{2008MNRAS.391.1137L,2010MNRAS.404..575L,2010MNRAS.404..590L,2019ApJ...872...76N}. Regardless of any added ingredients to the synthetic images (sky noise, resolution degradation, additional sources, etc.), the simulations provide foreknowledge of the true properties of each target: merger stage, mass ratio, gas fractions, initial morphologies, orbital parameters, etc. Consequently, one always has unbiased target classes upon which to evaluate the performance of the method. Furthermore, one can measure the biases affecting performance from both merger and image properties. Lastly, since the morphological features of galaxy interactions are induced primarily via gravitational effects, they should be largely insensitive to the particularities of the hydrodynamic model.

Two classes of methods which have gained significant traction in general astronomy and, in particular, galaxy astronomy are machine learning and deep learning (e.g. \citealt{2014MNRAS.439.3526T,2017Natur.548..555H,2019MNRAS.485..666B,2019NatAs...3...93R,2019MNRAS.484.5330J,2019MNRAS.486.3702S,2019ApJ...876...82N,2019arXiv190611248H}). Specifically, convolutional neural networks (CNNs) have been used to improve automated image-based galaxy morphology classifications with great success \citep{2015ApJS..221....8H,2018MNRAS.476.3661D}. The level of intricacy in the features that can be identified by CNN and other machine learning models has made them an attractive tool for merger identification (e.g. \citealt{2018MNRAS.479..415A,2019MNRAS.483.2968W}). Following the approaches adopted for quantitative morphologies by calibrating on hydrodynamical simulations, \cite{2019A&A...626A..49P} train a CNN on synthetic Sloan Digital Sky Survey (SDSS) images of galaxies from the EAGLE simulation \citep{2015MNRAS.446..521S} and examine biases from redshift, star-formation rates, and apparent brightness on merger and non-merger classifications -- though with poor classification performance ($65.5\%$ in a binary classification). Nonetheless, one of the key elements of their synthetic SDSS images is that they were inserted into a handful of SDSS survey fields in an attempt to match observational biases in real images (realistic skies, resolution, and crowding by nearby sources). Indeed, \cite{2019MNRAS.489.1859H} used a similar but more rigorous approach with CNNs trained on the \cite{2010ApJS..186..427N} SDSS visual classification sample to perform Hubble type classifications of synthetic images of galaxies from the Illustris\textsc{TNG-100} simulations \citep{2018MNRAS.475..648P,2018MNRAS.475..624N,2019MNRAS.483.4140R}. \cite{2019MNRAS.489.1859H} found that injecting the TNG images into real fields following the statistical observational realism approach of \cite{2017MNRAS.467.1033B} was crucial to obtaining consistent classification uncertainties when testing on SDSS and \textsc{TNG} images.

These previous studies touch upon core unanswered questions for training deep neural networks based on hydrodynamical simulations (and that are particularly relevant for characterizing merger stage). Namely, what kind of synthetic images should be used when training using simulations? How realistic do the images have to be in order to achieve high performance in identifying and characterizing mergers by stage in real images? Does a network that is trained on images which include contaminating effects (such as realistic skies, resolution degradation and additional sources in the image field-of-view (FOV)) perform better when handling new data which also contain these contaminants? In other words, what is gained by making synthetic images more realistic? Dust-inclusive radiative transfer can be used to generate photo-realistic images (e.g. \citealt{2006MNRAS.372....2J,2010MNRAS.403...17J,2011ApJS..196...22B,2015A&C.....9...20C}) but it is costly from computational and data storage perspectives. Is radiative transfer essential to merger classifications or can it be replaced with simpler images? These questions come at an important time when the state-of-the-art cosmological hydrodynamical simulations produce realistic and statistically representative populations of galaxies (e.g. IllustrisTNG; \citealt{2018MNRAS.475..648P,2018MNRAS.475..624N}). Crucially, mergers identified from these simulations' merger trees cover a range of mass ratios, orbital parameters, and initial galaxy properties that are comparable to the real Universe. Consequently, synthetic images generated along each merger sequence can be used to generate and calibrate deep network models to identify and characterize mergers in current and next-generation observational imaging surveys. 

The goals of this paper are to: (1) provide the methodology with which CNNs, trained and calibrated using hydrodynamical simulations, can be used to identify mergers and predict merger stage in realistic images and (2) assess the importance of realism in the synthetic training images.\footnote{It should be noted that, based on the results of \cite{2019MNRAS.489.1859H} who use our methods for Hubble type classifications, the applications of our methods and results are not restricted to mergers.} To realize these goals, we construct synthetic images with various levels of observational realism from a set of binary hydrodynamical merger simulations run with the \textsc{FIRE-2} model \citep{2019MNRAS.485.1320M}. Specifically, we generate images in two branches, starting with: (a) 2-D projections of the stellar particles and (b) photometry from dust-inclusive \textsc{skirt} radiative transfer. In each branch, images are are constructed with three levels of realism: (i) no observational effects (idealized); (ii) realistic skies and point-spread functions (semi-realistic); and (iii) statistical insertion into real survey images (fully realistic). These levels are designed to expose the roles of particular ingredients of realism in classification performance. Each image is assigned a target classification (isolated, pair or post-merger) corresponding to the definitions described in Section \ref{sec:selection} and illustrated in Figure \ref{fig:selection}. Given that all training, validation, and test data are drawn from the same set of isolated/merger simulation runs, the training data are (by construction) highly generalizable to the test data in terms of the range of galaxy/merger properties covered. This experimental design allows us to isolate the role of realism in the performances of the networks. 

This paper is laid out as follows. The simulations, construction of the synthetic images, and neural network architecture are described in Section \ref{sec:methods}. Our experiments and their results are presented in Section \ref{sec:experiments}. Our results are discussed in Section \ref{sec:discussion} and summarized in Section \ref{sec:summary}. We adopt a cosmology in which ($H_0 = 70$ km s$^{-1}$ Mpc$^{-1}$, $\Omega_{\mathrm{m}} = 0.3$, $\Omega_{\Lambda}=0.7$). Additionally, the \cite{2017MNRAS.467.1033B} observational realism suite, \textsc{RealSim}, is released publicly as a companion to this paper (see Section \ref{sec:SMFR}).




\section{Methods}\label{sec:methods}

In this section, we describe the merger simulations (Sections \ref{sec:model} and \ref{sec:suite}), merger stage definitions and snapshot selection (Section \ref{sec:selection}), creation of the synthetic images (Section \ref{sec:types}) and CNN architecture (Section \ref{sec:networks}).


\subsection{Simulations}\label{sec:simulations}

We use the suite of galaxy interaction simulations from \cite{2019MNRAS.485.1320M} in this study. The suite is similar to a previous merger suite from those authors \citep{2013MNRAS.433L..59P,2015MNRAS.448.1107M} but with much higher resolution and a new physical model and hydrodynamic solver. We describe the salient features of the suite here but refer the reader to \cite{2019MNRAS.485.1320M} and \cite{2018MNRAS.480..800H} for full details of the suite and model, respectively.\footnote{Videos of the \citet{2019MNRAS.485.1320M} galaxy merger simulations are available at \href{research.pomona.edu/galaxymergers}{research.pomona.edu/galaxymergers}.} We discuss the limitations with respect to the scope of the merger suite in detail in Section \ref{sec:limitations}. Briefly, we emphasize that the suite does not offer sufficiently representative statistics and diversity in galaxy/merger properties to train networks that will be useful in applications to a real population of galaxies. Therefore, we do not apply our trained networks to real galaxies. However, for the objectives highlighted at the end of Section \ref{sec:intro}, the suite is appropriate.

\subsubsection{\textsc{FIRE-2} model}\label{sec:model}
The simulations were run using the \textsc{FIRE-2} physics model \citep{2018MNRAS.480..800H} and the ``meshless finite-mass'' (MFM) hydrodynamics solver, \textsc{gizmo} \citep{2015MNRAS.450...53H,2017arXiv171201294H}.\footnote{For more information on the \textsc{FIRE} Project and \textsc{FIRE-2}, visit \href{https://fire.northwestern.edu}{https://fire.northwestern.edu}.} The model includes treatment of radiative cooling and heating from free-free, photo-ionization and recombination, Compton, photoelectric, dust-collisional, cosmic ray, molecular, metal-line and fine-structure processes. It accounts for the UV background \citep{2009ApJ...703.1416F} and locally-driven heating and self-shielding. Gas which is locally self-gravitating, self-shielding, Jeans unstable and sufficiently dense (defined by critical gas density, $n_{\mathrm{crit}}=1000$ cm$^{-3}$) can form stars stochastically in a sink-particle approach (see Appendix C of \citealt{2018MNRAS.480..800H}). A stellar particle is treated as a single stellar population with a known age, $t_{\star} = t-t_{\mathrm{form}}$, and a metallicity and mass which are inherited from its progenitor gas particle. Masses, ages, metalliticities, luminosities, energies, mass-loss rates and stellar feedback event rates are tabulated (without tuning) using the \textsc{Starburst99} stellar population synthesis model \citep{1999ApJS..123....3L} assuming a \cite{2001MNRAS.322..231K} initial mass function (IMF). Stellar feedback includes: (i) mass, metal, energy, and momentum injection from supernova type Ia \& II; (ii) continuous stellar mass-loss through OB/AGB winds; (iii) photo-ionization and photo-electric heating; and (iv) radiation pressure. The model does not account for feedback generated via accretion of gas onto supermassive black holes (SMBHs). SMBH feedback is omitted because coupling between an active galactic nucleus (AGN) and the circumnuclear interstellar medium (ISM) is not yet well understood (though see \citealt{2017MNRAS.467.2301T} for an examination of the stability of feedback regulated star-formation in galactic nuclei). The MFM dark matter, gas, and stellar particle masses are $(m_{\mathrm{dm}},m_{\mathrm{gas}},m_{\mathrm{star}})=(19,1.4,0.84)\times10^4\;M_{\odot}$. The highest gas density and spatial resolution are $5.8\times10^5$ cm$^{-3}$ and 1.1 pc, respectively. The typical snapshot resolution is 5 Myr. The gravitational softening lengths are 10 pc for dark matter and stellar components and 1 pc for the gaseous component.

\subsubsection{Merger suite}\label{sec:suite}

\begin{figure}
	\includegraphics[width=\linewidth]{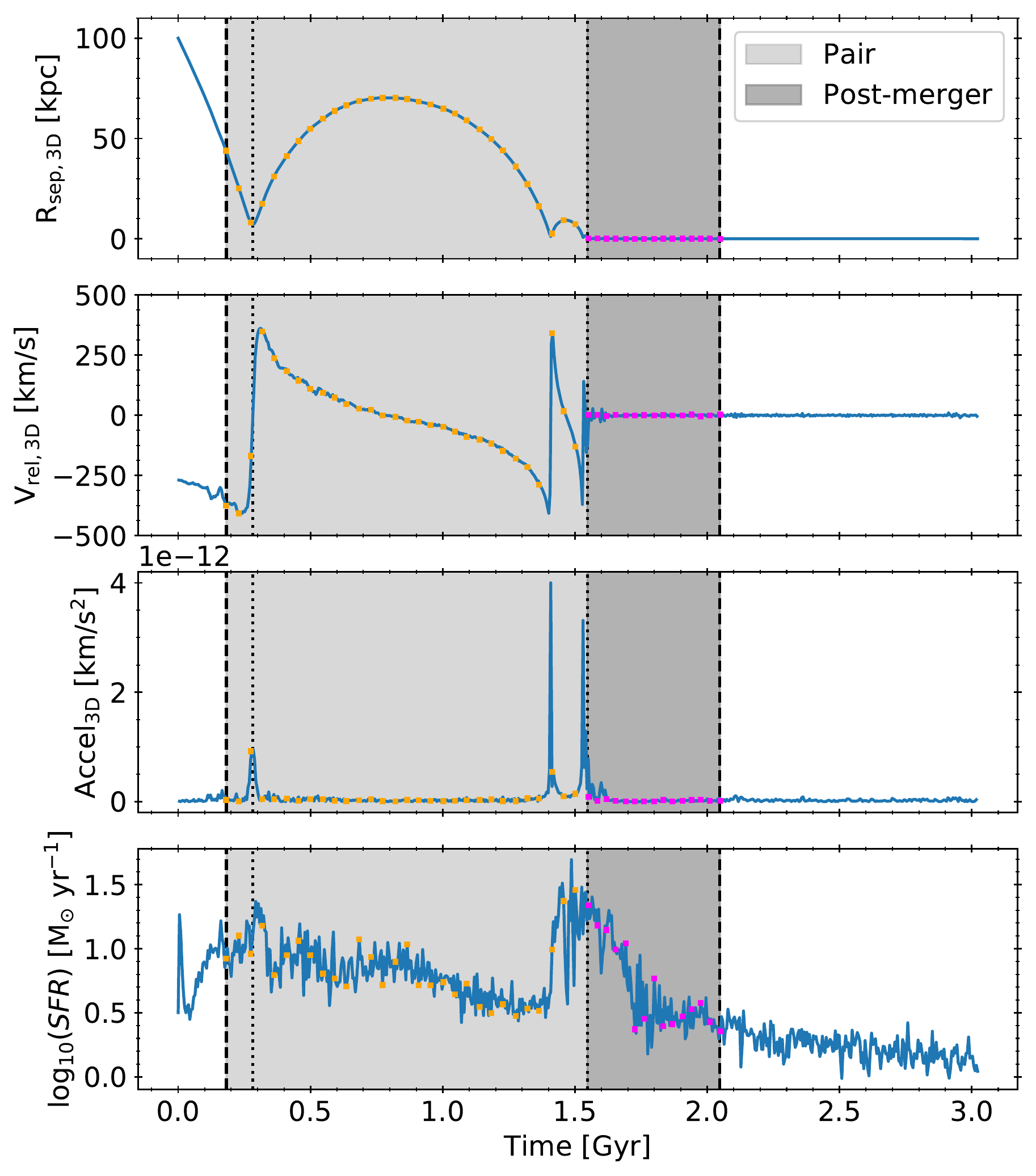}
    \caption{Radial separation, relative velocity, absolute acceleration, and total star formation rate (SFR) sequences for the G2G3 ``e'' orbit 1 merger (the fiducial run from \citealt{2019MNRAS.485.1320M}) showing class definitions and snapshot selection. From left to right in each panel, the first thick dashed and dotted lines correspond to 100 Myr before first pericentric passage and the moment of first pericentric passage, respectively. The second dotted line corresponds to coalescence -- which we take to be the last time the central black holes of each galaxy are more than 500 pc apart. The second thick dashed line corresponds to 500 Myr after coalescence. Shading between lines correspond to the pair (light gray) and post-merger (darker gray) classes. In each panel, we show the snapshot selection for the pair (orange) and post-merger (magenta) classes. The SFRs are measured from the full simulation volume and so include contributions from both galaxies when they are separate. }
    \label{fig:selection}
\end{figure}

\cite{2019MNRAS.485.1320M} used \textsc{FIRE-2} physics to generate a suite of non-cosmological binary galaxy interaction simulations covering a range of orbital parameters and mass ratios between four disc galaxies (G1, G2, G3, and G4, in order of increasing total stellar mass). The suite is complimented by secular runs (the controls used by \citealt{2019MNRAS.485.1320M}) in which each galaxy is allowed to evolve in isolation. Individual galaxies are set up following the procedure described in \cite{2005MNRAS.361..776S} using the analytic framework provided by \cite{1998MNRAS.295..319M}. Stellar bulges and dark matter haloes are initialized analytically with \cite{1990ApJ...356..359H} profiles. Halo masses are adopted for a given stellar mass following the abundance matching results from \cite{2013MNRAS.428.3121M}. Stellar bulge-to-total fractions are assigned on the basis of median trends with total stellar mass using the \cite{2014ApJS..210....3M} estimates of bulge, disc and total stellar masses for galaxies in the SDSS. Similarly, gas fractions are assigned based on mean atomic and molecular gas mass fractions estimates along the main sequence (MS) of star-forming galaxies in the SFR-$M_{\star}$ plane from \cite{2016MNRAS.462.1749S}. The properties of each galaxy are shown in Table \ref{tab:galaxy_properties}. 

\begin{table*}
	\centering
	\caption{Initial properties of the four galaxies in the \citealt{2019MNRAS.485.1320M} merger suite. The columns are the galaxy ID, total stellar mass, halo mass, gas fraction, stellar bulge-to-total mass fraction, gas disc scale length, and stellar disc scale length.}
	\label{tab:galaxy_properties}
	\begin{tabular*}{\linewidth}{@{\extracolsep{\fill}}lcccccccr} 
		\hline
		Galaxy ID & $M_{\star}/10^{10}M_{\odot}$ &  $M_{\star}/M_{\mathrm{halo}}$  & $f_{\mathrm{gas}}$ & $(B/T)_{\star}$ & $R_{\mathrm{d,gas}}$/kpc & $R_{\mathrm{d,stars}}$/kpc  \\
		\hline 
		G1 & 0.206 & 0.0157 & 0.681 & 0.0185 & 4.73 & 1.42  \\
		G2 & 1.24 & 0.0361 & 0.392 & 0.0497 & 6.04 & 1.92 \\
		G3 & 2.97 & 0.0383 & 0.264 & 0.0773 & 5.32 & 1.61  \\
		G4 & 5.50 & 0.0228 & 0.192 & 0.103 & 5.26 & 1.57 \\
		\hline
	\end{tabular*}
\end{table*}

The suite is divided into two components: (1) an orbit suite and (2) a mass ratio suite. The orbit suite comprises several interaction scenarios between the G2 and G3 galaxies (see Figure 3 of \citealt{2019MNRAS.485.1320M}). It covers three unique spin-orbit orientations corresponding to the ``e'', ``f'' and ``k'' orbits from \cite{2006ApJ...645..986R} (see Figure 1 of \citealt{2015MNRAS.448.1107M}), three impact parameters, and three impact velocities (see Figure 3 of \citealt{2019MNRAS.485.1320M}). The ``e'', ``f'' and ``k'' spin-orbit orientations correspond to approximately prograde, polar, and retrograde orbits, respectively. Permutation of the orbital parameters gives a total of $3\times3\times3=27$ unique mergers at a fixed mass ratio of $\mu\sim2.5:1$. The mass suite adopts a single orbit (the fiducial "e" orbit in \citealt{2019MNRAS.485.1320M}) in which each of the four galaxies interacts with every other galaxy and itself for a total of $4\times4=16$ interactions. The range in stellar mass ratios covered by the suite is $1:1$ to $1:16$. Combined, the orbit and mass components of the suite cover a broad range of encounter strengths and merging timescales. 

\subsubsection{Class definitions, merger selection, and snapshot sampling}\label{sec:selection}
For each merger simulation, we select snapshots from which to construct images based on a set of simple definitions of the pair and post-merger phases. For example, Figure \ref{fig:selection} shows the radial separation, relative velocity, absolute acceleration, and total star formation rate (SFR) sequences for the fiducial merger simulation from \cite{2019MNRAS.485.1320M} measured from central SMBHs of each galaxy. The pair phase is defined to begin 100 Myr before first pericentric passage, $t_p-100$ Myr, and to end just before coalescence, $t_c$ -- which, as in \cite{2019MNRAS.485.1320M}, we define as the last time the central black holes are more than 500 pc apart (light gray shading). We then define the post-merger phase as any time in [$t_c,t_c + 500$ Myr] (darker gray shading). 

We use a temporal definition for the beginning of the pair phase primarily for convenience with the simulations. But additionally, a temporal definition may circumvent biases that can arise from selection based on projected separation or relative velocity. For example, a 300 km/s relative velocity cut \citep{2008AJ....135.1877E,2013MNRAS.433L..59P} may have missed snapshots around first and second pericentre for the merger in Figure \ref{fig:selection} depending on line-of-sight.

Likewise, our definition of the post-merger phase is motivated by simplicity and the availability of temporal information in the simulations. Observability timescales for post-merger features (shells, streams, etc.), as for the mergers overall, are sensitive to surface brightnesses, gas fractions, mass ratios, and initial orbital parameters (e.g. \citealt{2008MNRAS.391.1137L,2010MNRAS.404..575L,2010MNRAS.404..590L,2014A&A...566A..97J,2019ApJ...872...76N}). Consequently, we select a clear-cut post-merger phase in which post-merger features are still expected to be prominent. Several mergers in the suite are either fly-bys or do not evolve to 500 Myr after coalescence. In order to preserve several mergers that coalesce but do not have 500 Myr of snapshot coverage after coalescence, we set a minimum post-coalescence criterion of 250 Myr. Consequently, the post-merger stage is defined as starting at coalescence and ending at $\max(t_{\mathrm{last}},t_c+500$ Myr$)$ where $t_{\mathrm{last}}\geq t_c+250$ Myr. The criteria that (a) the galaxy must coalesce and (b) the simulation has run for at least 250 Myr post-coalescence reduces the sample from 43 to 23 mergers. In particular, the interactions with the largest mass ratios and widest impact parameters are rejected on the basis of these criteria. We discuss the consequences of our class definitions on network performance in Section \ref{sec:disc_classes}.

For each merger simulation, 30 (15) snapshots are selected which uniformly sample the pair (post-merger) phase as shown with the orange (magenta) squares in Figure \ref{fig:selection}. This approach provides a sampling cadence that is sparse enough that imaging in neighbouring snapshots are not overly correlated (see Appendix \ref{sec:corr}) yet fine enough that a large number of images for each merger can be generated. Ten snapshots are selected from the isolated runs with uniform sampling cadence as with the pairs and post-mergers. Due to the significantly smaller number of snapshots corresponding to isolated galaxy runs, we bolster the isolated galaxy dataset by increasing the number of orientations in which their synthetic images are generated (as described below).

\subsection{Synthetic Observations}\label{sec:types}

Synthetic images are made for isolated, pair and post-merger snapshots along four lines of sight corresponding to the arms of a tetrahedron whose vertex is coincident with the point of minimum potential. Consequently, images in the pair phase are always centred on the more massive galaxy.\footnote{For our purposes, this bias toward the more massive galaxy is not of any consequence. But for a dataset which must be large and general enough to handle real data, separate images should be constructed that are centred on both the primary and secondary -- ideally in along different lines-of-sight.} Producing only 4 camera angles for the isolated galaxies would make the training data highly imbalanced -- with an order of magnitude more images with pair and post-merger classes than the isolated class.\footnote{Class imbalance -- where the occurrence of one class in a dataset significantly outnumbers other classes -- is a common problem in applications of deep learning to classification tasks including medical diagnosis \citep{grzymala2004approach,mac2002problem}, fraud detection \citep{chan1998toward} and others \citep{radivojac2004classification,cardie1997improving,haixiang2017learning}. An example is the natural imbalance in medical data between images of a particular diagnostic class (e.g. contains tumour) which might be 1000 times less frequent than images of another class (e.g. healthy). Unbalanced data can have significant detrimental effects on deep classifiers such as CNN and a recent systematic study has investigated various methods which address class imbalance \citep{2017arXiv171005381B}. Indeed, the method that appears to best address class imbalance is oversampling of data from the under-represented class (e.g. through augmentation) -- which is the method we adopt in this paper.} As a first step in balancing the data, we increase the number of camera angle orientations for snapshots from the isolated runs by 11 inclinations and 11 position angles. Before any augmentation (see Section \ref{sec:normalization}), there are $23\times30\times4=2\,760$ images with the pair class, $23\times15\times4=1\,380$ with the post-merger class, and $4\times10\times(11\times11+4)=5\,000$ with the isolated class. 

Synthetic images are generated for each snapshot/orientation with various levels of realism. There are two distinct image types: (1) images originating from two-dimensional projections of stellar particles (stellar maps, \textsc{Sm}) and (2) from photometry generated using dust-inclusive radiative transfer (\textsc{Ph}). We produce images with three different levels of realism: (1) noiseless with high resolution; (2) include realistic (but analytically generated) noise and resolution degradation; and (3) are inserted into real SDSS survey fields that may contain additional sources. We refer to these increasing levels of realism as ``idealized'', ``semi-real'' and ``full real'', and they allow us to examine the importance of observational biases that are introduced level-by-level. The image types are described in detail in the sections that follow and are summarized in Table \ref{tab:types}.


\subsubsection{\textbf{StellarMap} (\textsc{Sm})}\label{sec:SM}
The zeroth order stellar image that can be produced from a hydrodynamical simulation is a two-dimensional projection of the stellar particles along a given line-of-sight. This \emph{idealized} image type has several important features: noiseless without resolution degradation; insensitivity to variations in mass-to-light ratio ($M/L$) from different stellar populations or dust absorption; and low computational and data management overhead. Initially, we adopt a fixed 50 kpc field of view (FOV) for each image with spacial resolution of 0.097 kpc/pixel ($512\times512$ pixels).\footnote{The higher the initial resolution, the greater the range of instrumental angular scales (arcsec/pixel) and redshifts (kpc/arcsec) that can be explored.} All images (including other types) are mock-observed with the SDSS camera (0.396 arcsec/pixel) at a fixed redshift of $z=0.046$ (the median redshift of galaxies in the DR14 MaNGA galaxy sample \citealt{2015ApJ...798....7B}) where the scale is approximately 0.9 kpc/arcsec. Consequently, the \textsc{Sm} images are re-binned to a physical scale of 0.36 kpc/pixel ($139\times139$ pixels or $56\times56$ arcsec$^2$). This still offers high resolution -- particularly with respect to images that are further degraded by realistic (or real) SDSS Point-Spread Functions (PSFs). 

\subsubsection{\textbf{Photometry} (\textsc{Ph})}\label{sec:Ph}
We generate idealized SDSS $gri$ photometric images using the Monte Carlo dust radiative transfer code, \textsc{skirt} \citep{2011ApJS..196...22B,2015A&C.....9...20C}. \textsc{skirt} predicts the light contribution from stellar particles and star-forming regions whilst modelling the effects of dust on the absorption, scattering, and re-emission of stellar light (note that we ignore radiation from the central engine). We model the stellar light from old stars (older than 10 Myr) using a \cite{2001MNRAS.322..231K} initial mass function and the associated \textsc{starburst99} single-age spectral energy distributions (SEDs) \citep{1999ApJS..123....3L}. Emission from star-forming regions (stellar particles younger than 10 Myr) is represented by \textsc{mappings-iii} SEDs \citep{2008ApJS..176..438G} which include contributions from young stars and \ion{H}{ii} regions. The dust contribution is modelled assuming that the dust distribution traces the metal distribution where $30\%$ of the metals are locked in dust particles. We adopt the multi-component dust mix of \cite{2004ApJS..152..211Z} which includes graphite grains, silicate grains, and polycyclic aromatic hydrocarbons (PAHs). We ignore dust re-emission (and the associated self-absorption) which has a negligible contribution in the wavelength regime studied in this work. The underlying radiation field is discretized by \textsc{skirt} using $10^5$ photon packages per wavelength. \textsc{skirt}'s output (spectral datacubes) are converged for $>10^5$ photon packages per wavelength.

We adopt the same initial FOV and spatial resolution as for the \textsc{Sm} images in construction of the \textsc{skirt} datacubes. Since we are generating broadband photometry, a relatively coarse spectral resolution is adopted with 241 spectral elements which linearly and uniformly sample the rest-frame optical spectrum from the near-UV to near-infrared ($250-850$ \AA). These datacubes are redshifted to $z=0.046$ and convolved with the SDSS $gri$ response functions to produce idealized photometry --  accounting for stretch in the spectrum and $(1+z)^{-5}$ reduction in specific intensities in each spectral element.\footnote{We provide a code which performs all of these tasks (for a specified redshift) from the default rest-frame specific intensity datacubes from SKIRT (W m$^{-2}\mu$m$^{-1}$ arcsec$^{-2}$). The code produces output photometry in convenient AB mag/arcsec$^2$ units for each filter \citep{1983ApJ...266..713O} and can be found at the following url: \href{https://github.com/cbottrell/RealSim/blob/master/SpecToSDSS_gri.py}{https://github.com/cbottrell/RealSim/blob/master/SpecToSDSS\_gri.py}.} As with the \textsc{Sm} images, the \textsc{Ph} images are rebinned to the SDSS camera pixel scale. The \textsc{Ph} images are: noiseless with high resolution; light-weighted and sensitive to variations in $M/L$ for different stellar populations and dust; and very expensive from a computational and data management perspective when compared to \textsc{Sm} (see Section \ref{sec:SMvsPh}). Furthermore, training networks with all three $gri$ bands as input allows networks to develop sensitivity to colour. 

\begin{table*}
	\centering
	\caption{Reference summary of image types used for training and testing of networks. Intensities in the \textsc{StellarMap} images are scaled to match the total surface brightnesses in the \textsc{Photometry} $i$-band images before adding realism effects. }
	\label{tab:types}
	\begin{tabular*}{\linewidth}{@{\extracolsep{\fill}}lcccccccr} 
		\hline
		Image Type & Shortform &  Radiative Transfer & Bands & Gaussian Sky & Gaussian PSF & Real Sky & Real PSF  \\
		\hline 
		\textsc{StellarMap} & \textsc{Sm} & no & $i^{\star}$ & no & no & no & no \\
		\textsc{StellarMap SemiReal} & \textsc{SmSR} & no & $i^{\star}$ & yes & yes & no & no \\
		\textsc{StellarMap FullReal} & \textsc{SmFR} & no & $i^{\star}$ & no & no & yes & yes \\
		\textsc{Photometry} & \textsc{Ph} & yes & $gri$ & no & no & no & no  \\
		\textsc{Photometry SemiReal} & \textsc{PhSR} & yes & $gri$ & yes & yes & no & no \\
		\textsc{Photometry FullReal} & \textsc{PhFR} & yes & $gri$ & no & no & yes & yes \\
		\hline
	\end{tabular*}
\end{table*}

\subsubsection{\textbf{StellarMap SemiReal} (\textsc{SmSR})}\label{sec:SMSR}

\begin{table*}
	\centering
	\caption{SDSS sky and angular resolution measurements used to generate the background noise levels and convolution kernels for SemiReal images. Table quantities are computed from the ensemble of SDSS Field table values corresponding to the \emph{full} Simard et al. (2011) galaxy sample. Columns: (1) SDSS bandpass; (2) Mean sky noise [AB mag/arcsec$^2$]; (3) Standard deviation in sky noise values [AB mag/arcsec$^2$]; (4) Mean PSF FWHM [arcsec]; (4) Standard deviation in PSF FWHM values [arcsec]. Individual SemiReal sky noise values and PSFs are drawn from normal distributions formed from these quantities. }
	\label{tab:SemiReal}
	\begin{tabular*}{\linewidth}{@{\extracolsep{\fill}}ccccc} 
		\hline
		SDSS band & $\langle \sigma_{\mathrm{sky,Field}} \rangle$ [AB mag/arcsec$^2$] & stdev($\sigma_{\mathrm{sky,Field}}$) [AB mag/arcsec$^2$] & $\langle$ FWHM$_{\mathrm{PSF}}$ $\rangle$ [arcsec] & stdev(FWHM$_{\mathrm{PSF}}$) [arcsec] \\
		\hline
		u & 23.87 & 0.15 & 1.55 & 0.24\\
		g & 24.88 & 0.14 & 1.47 & 0.22\\
		r & 24.38 & 0.11 & 1.36 & 0.22\\
		i & 23.82 & 0.12 & 1.29 & 0.22\\
		z & 22.36 & 0.19 & 1.31 & 0.20\\
		\hline
	\end{tabular*}
\end{table*}

Ground-based imaging surveys are affected by sky surface-brightness limitations and blurring from the atmospheric PSF. These biases can be emulated using the statistics of sky brightnesses and PSF sizes measured in SDSS fields. Crucially, we match the \emph{statistics} of sky noise levels and PSF resolution to the field properties for 1.12 million galaxies in the SDSS Legacy images \citep{2009ApJS..182..543A} using the \cite{2011ApJS..196...11S} quantitative morphology catalog and ancillary data measured by the \texttt{PHOTO} pipeline \citep{2001ASPC..238..269L,2002SPIE.4836..350L,2012photolite}. We compute the means and standard deviations in the resulting sky noise and PSF resolution distribution functions. The results are tabulated in Table \ref{tab:SemiReal}. We use these results to generate analytic Gaussian profiles from which sky noise and PSF resolution levels are sampled independently for each synthetic image. 

The idealized \textsc{Sm} images are not light-weighted and therefore do not offer straight-forward conversion to calibrated AB flux units. To approximate the intensities of the stellar maps in realistic images, we scale each normalized stellar map by the total intensity in its corresponding idealized $i$-band photometry image. We choose to scale by the $i$-band light because it is less sensitive to variations in $M/L$ from young stellar populations or starbursts compared to $g$ or $r$. With the idealized \textsc{Sm} images effectively ``light-weighted'', we sample the distribution function for the PSF and convolve. Before adding sky noise, we use the average SDSS photometric zero-point magnitude, airmass, extinction, and gain over all SDSS fields to convert the PSF-convolved images to electron counts from which source Poisson noise can be added. We then convert back to calibrated flux units, sample our sky noise distribution and add Gaussian sky noise to the image. 

\subsubsection{\textbf{Photometry SemiReal} (\textsc{PhSR})}\label{sec:PhSR}

The procedure for creating \textsc{PhSR} images in each band is the same as for creating \textsc{SmSR} images but without the normalization and ``light-weighting'' step. One feature of the current \textsc{SemiReal} procedure that can be remedied in the future is that the sky noise and PSF estimates are drawn independently in each band and so are not correlated as they should be. However, our results do not give us reason to suspect that this is a significant limitation of our methods.

\subsubsection{\textbf{StellarMap FullReal} (\textsc{SmFR})}\label{sec:SMFR}

Synthetic images with extensive observational realism are generated following the methods presented in \cite{2017MNRAS.467.1033B,2017MNRAS.467.2879B}. Similar to \textsc{SemiReal}, the \textsc{FullReal} procedure is designed to incorporate \emph{statistical} observational realism into the synthetic images so that real survey field statistics are matched between the simulated and observed galaxies. The main difference from the \textsc{SemiReal} procedure is that the synthetic images are added quasi-randomly to real survey fields in the \textsc{FullReal} procedure. In this approach, the insertion statistics are guided by a basis catalog of real galaxies \citep{2011ApJS..196...11S}. As such, the statistics of sky brightness, PSF resolution and crowding by nearby sources for real galaxies are reproduced in the synthetic images (along with any other field-dependent characteristics). The \textsc{FullReal} procedure is described in detail in \cite{2017MNRAS.467.1033B}. We provide a summary here of the procedure that is followed for every synthetic image to compliment the public release of the realism suite\footnote{A public version of the \textsc{RealSim} observational realism suite is available the following url: \href{https://github.com/cbottrell/RealSim}{https://github.com/cbottrell/RealSim} for Python 3. It includes: the \cite{2011ApJS..196...11S} bulge+disc decomposition catalog from which it draws galaxy and field statistics; a Python 3 version of the SDSS \texttt{sqlcl.py} code which queries field information directly from the SDSS Data Archive Server; $ugriz$ filter response functions from \cite{2010AJ....139.1628D}; the \cite{2011ApJS..196...11S} \textsc{SExtractor} configuration files required for deblending of images when inserting into real SDSS fields; a Python notebook of example executions; a code for converting \textsc{skirt} output to SDSS images; and a sample of input images.}:
\begin{enumerate}
\item A galaxy is randomly selected from the \cite{2011ApJS..196...11S} basis catalog. The SDSS $gri$-band fields in which that galaxy resides are extracted and converted to calibrated flux units using ancillary data queried from the SDSS Data Archive Server. A source mask is generated for the $r$-band field using \textsc{SExtractor} \citep{1996A&AS..117..393B} and deblending parameters optimized for SDSS in \cite{2011ApJS..196...11S} (specifically for the \citealt{2011MNRAS.412..591P} pair sample). A common injection site for each band (where applicable) is selected randomly with the restriction that the \emph{centre} of the injected image cannot land on another object in the source mask.
\item The PSFs for each band corresponding to the injection site are reconstructed using the \textsc{psField} files and the standalone \href{https://www.sdss.org/dr15/algorithms/read_psf/}{\textsc{read\_psf}} software. Each band of the synthetic image (in the \textsc{SmFR} case, the single ``light-weighted'' idealized stellar map) is converted to electrons using the ancillary data from Step (i) and convolved with the local SDSS PSF for that band. Source Poisson noise is then added. 
\item A PSF-convolved and Poisson noise-added synthetic image (in each desired band) is finally converted back to calibrated flux units and inserted into the SDSS field at the injection site selected in Step (i). A cutout corresponding to the desired FOV (in our case, 50 kpc or approximately 56 arcsec at $z=0.046$) is extracted around this location. This cutout now includes real sky, real PSF degradation and real additional sources in the FOV which track the statistics for galaxy image properties in the basis catalog.
\end{enumerate}

Some particularities of this procedure for generating \textsc{SmFR} images are that (1) images are only generated in the $i$-band and (2) the $r$-band image is still used to generate the source mask as described in Step (i).

\subsubsection{\textbf{Photometry FullReal} (\textsc{PhFR})}\label{sec:PhFR}
Construction of \textsc{PhFR} images follows the same procedure as for the \textsc{SmFR} but for each of the SDSS $gri$ bands. Because the \textsc{PhFR} images incorporate light-weighting from radiative transfer and the full rigour of statistical observational realism, the \textsc{PhFR} dataset is our benchmark for how a given network would be expected to perform on realistic data and is often discussed as such in Sections \ref{sec:experiments} and \ref{sec:discussion} (i.e. the \textsc{PhFR} images are the closest representation of observable galaxies in our suite and are hence used as the ultimate training set for likely "real life" performance). As with the \textsc{Ph} and \textsc{PhSR} datasets, the \textsc{PhFR} has three channels of input corresponding to the three bands in which we produce photometry.\\

\begin{figure*}
	\includegraphics[width=\linewidth]{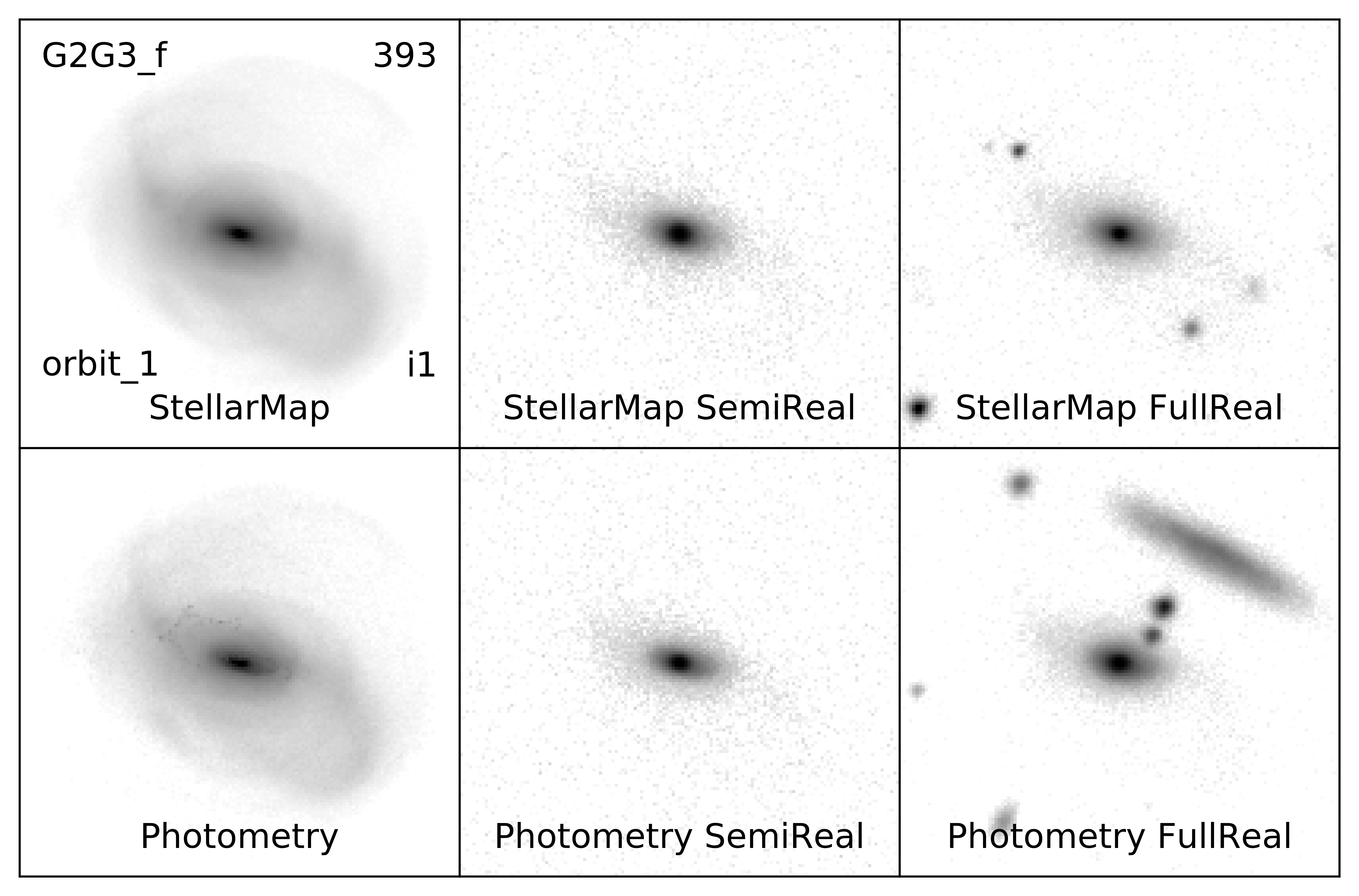}
    \caption{Visualization of a single post-merger galaxy realized with every image type. The post-merger image taken from orbit 1 of the G2G3 ``f'' orbit suite at snapshot 393 (post-coalescence). All images show $i-$band intensities. Note the potential for misclassification in the \textsc{PhFR} realization of this post-merger due to a chance projection with a field galaxy.}
    \label{fig:types}
\end{figure*}

Figure \ref{fig:types} shows a recent post-merger for each of the 6 image types and demonstrates the impact of each level of realism. The upper panels show images originating from stellar maps and lower panels show images originating from radiative transfer. In the idealized \textsc{Sm} and \textsc{Ph} images (left hand column), morphological features post-merger are visually prominent including shells, streams, and tidal tails that have not yet decayed from the pair phase. In the \textsc{Ph} image, a dust lane obscures light emanating from the nucleus -- giving it an asymmetric appearance with respect to the \textsc{Sm} image. Additionally, the \textsc{Ph} image has bright knots associated with the low $M/L$ of young stellar populations whereas the \textsc{Sm} image is insensitive to these features. The \textsc{SemiReal} images in the middle panels show the results of adding realistic SDSS noise and resolution degradation to the images. Many of the features that made this object easily identifiable as a post-merger in the idealized images are drowned by the sky noise and PSF blurring. Features of the post-merger remain in the \textsc{SemiReal} images but are more subtle than in idealized images. The right panels show \textsc{FullReal} images for the post-merger. In addition to real skies and degradation by real PSFs, these images incorporate contamination by nearby sources. The lower right panel shows particularly striking chance projection with an interloping disc galaxy. Taken together, these images nicely encapsulate the rationale of this work.

\subsection{Image normalization and augmentation}\label{sec:normalization}

Normalizations and augmentation (oversampling) are applied to images in each database. We generate augmented images by applying zoom, rotation, and small translational transformations to the set of original images in order to (1) reduce class imbalance and (2) achieve rotational invariance in the models. The augmentations are performed with the using the \textsc{ImageDataGenerator} class from the \textsc{Keras} Python API \citep{chollet2015keras}. All images of a particular class are augmented $N$ times until the total number of images exceeds 10,000. Consequently, the final number of images in each class are $(N_{\mathrm{Iso}},N_{\mathrm{Pair}},N_{\mathrm{Post}}) = (10000,11040,11040)$ after augmentation for each image type.

The augmented images (starting in linear intensities) are normalized following a standardized algorithm which is applied to all image types:
\begin{itemize}
\item If the image contains sky noise (\textsc{SemiReal} and \textsc{FullReal} types), then the image is subtracted by its median intensity.
\item Take the logarithm of the sky-subtracted image. All values less than -7 are converted to NaNs. 
\item The median of the full image, $a_{\mathrm{min}}$, and 99$^{\mathrm{th}}$ percentile for the central $20\times20$ pixels, $a_{\mathrm{max}}$, are computed.
\item All values below $a_{\mathrm{min}}$ (and NaNs) are set to $a_{\mathrm{min}}$ and values greater than $a_{\mathrm{max}}$ are set to $a_{\mathrm{max}}$.
\item The clipped logarithmic image is subtracted by $a_{\mathrm{min}}$ and normalized by $a_{\mathrm{max}} - a_{\mathrm{min}}$.
\end{itemize}

The results are images with logarithmic-scale intensities in the range of 0 to 1 in which each image is scaled to maximize contrast for the central target in the image. This normalization procedure avoids the problem of reduced contrast in \textsc{FullReal} images which contain bright stars or other contaminating effects (such as would be produced by a more conventional standard scalar). 

\subsection{Neural network architecture}\label{sec:networks}


\begin{table}
	\caption{Convolutional neural network architecture for the full-colour model which accepts 3-channels of input comprising $gri$ images. Convolution kernel sizes (Conv2D), max-pooling windows (MaxPool), dropout rates (Dropout), output shapes and total number of trainable parameters for each of the layers are indicated. The convolution layers use Rectified Linear Unit (ReLU) non-linear activation functions and have a (1,1) stride. The output of the 4th convolution layer is flattened to a one-dimensional feature array and passed to the fully connected (dense) component. Dense layers also use ReLU activation functions. The output layer uses a softmax activation function.}
	\label{tab:networks}
	\begin{center}
	\begin{tabular*}{\linewidth}{@{\extracolsep{\fill}}lcc} 
		\hline
		 Layer (type) & Output shape & \# Parameters  \\
		\hline \hline
		Input Layer & $(139,139,3)$ & 0\\
		\hline
		Conv2D-1 $(6\times6)$ & $(139,139,32)$ & 3\,488\\
		MaxPool-1 $(2\times2)$ & $(69,69,32)$ & 0\\
		Dropout-1 (0.25) & $(69,69,32)$ & 0\\
		\hline
		Conv2D-2 $(5\times5)$ & $(69,69,64)$ &  51\,264\\
		MaxPool-2 $(2\times2)$ & $(34,34,64)$ & 0\\
		Dropout-2 (0.25) & $(34,34,64)$ & 0\\
		\hline
		Conv2D-3 $(2\times2)$ & $(34,34,128)$ &  32\,896 \\
		MaxPool-3 $(2\times2)$ & $(17,17,128)$ & 0\\
		Dropout-3 (0.25) & $(17,17,17)$ & 0\\
		\hline
		Conv2D-4 $(3\times3)$ & $(17,17,128)$ & 147\,584\\
		Dropout-4 (0.25) & $(17,17,17)$ & 0\\
		Flatten & $36\,992$ & 0 \\
		\hline\hline
		Dense-1 & (512) & 18\,940\,416 \\
		DropFC-1 (0.25) & (512) & 0 \\
		\hline
		Dense-2 & (128) & 65\,664 \\
		DropFC-2 (0.25) & (128) & 0 \\
		\hline\hline
		Output Layer & (3) & 387\\
		\hline\hline
		Total \# parameters & & 19\,241\,699\\
		\hline
	\end{tabular*}
	\end{center}
\end{table}

We use our synthetic images to train CNNs which classify galaxies as isolated, pairs, or post-mergers. CNNs are a class of deep learning model that are particularly useful for data which exhibit topological structure such as images \citep{Fukushima1980,lecun1989backpropagation,lecun1995convolutional,lecun1998gradient,Krizhevsky:2012:ICD:2999134.2999257,2015Natur.521..436L}. There is enormous flexibility in CNN architectures in terms of depth (number of layers), layer properties (kernel sizes, etc.) and layer structures (e.g. residual blocks, \citealt{2015arXiv151203385H}). Given the successes of previous works using a particular (and relatively simple) CNN architecture for galaxy morphology classifications \citep{2015MNRAS.450.1441D,2015ApJS..221....8H,2018MNRAS.476.3661D,2019MNRAS.484...93D,2019MNRAS.489.1859H}, we adopt a similar (but not identical) CNN architecture for predicting galaxy merger stage. This architecture is summarized in Table \ref{tab:networks}. The output of the network for a given image is a class probability distribution function, $(P_{\mathrm{Iso}},P_{\mathrm{Pair}},P_{\mathrm{Post}})$. For our analyses and comparison with the known target classes, we adopt the class with the highest probability density.

A $(70,15,15)\%$ split is used for \emph{training}, \emph{validation}, and \emph{test} images. The networks are optimized on the training images and corresponding known target classes. The overall performance of a network on the validation images, $N($correct$)/N_{\mathrm{tot}}$, is evaluated after each training epoch. While this step does not, strictly-speaking, affect optimization, it is used to determine an appropriate time to stop training and consequently prevent overfitting to the training images. In contrast, networks are never exposed to test images during training. For tests in which networks trained on a particular image type (e.g. \textsc{PhFR}) are tested on images of a different type (e.g. \textsc{SmSR} -- which may \emph{all} technically be considered distinct data), we find that there is no difference in network performance whether we test only on the corresponding test images or all images (including training and validation images). Our results show the latter for such tests.

\section{Experiments}\label{sec:experiments}

Each synthetic image dataset is used to train neural networks which are then applied to test images of every type. This ``handshake'' of training/testing experiments includes cases where training and test data are of the same type. We generate 10 networks for each image type by splitting the data into training, validation and test images using 10 unique random states. This ``bootstrapping'' of the data allows us to characterize the random error associated with the selection of a particular training set and to statistically merge our test results.

\subsection{Model and image handshake}\label{sec:handshake}

\begin{figure}
	\includegraphics[width=\linewidth]{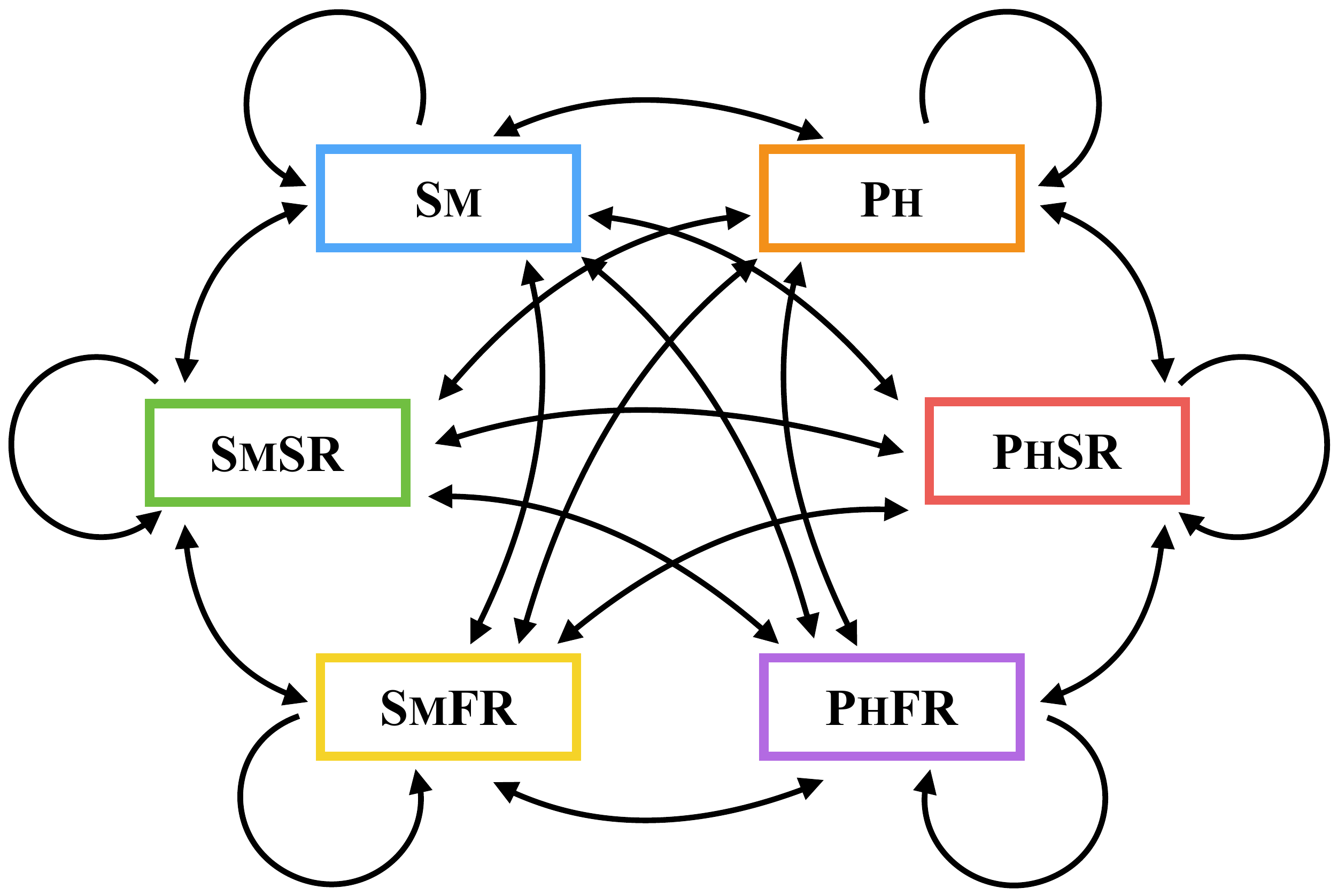}
    \caption{Schematic of the main handshake experiment. Networks are trained with images of each type. Each trained network is tested to images of every other type. Additionally, each network is tested on images of the same type but that the networks never see during training (outer looping arrows).}
    \label{fig:handshake}
\end{figure}

Our main experiment comprises a handshake between networks trained using the 6 unique image types (see Table \ref{tab:types}) which are each tested on the 6 image types -- resulting in $6\times6$ tests. Figure \ref{fig:handshake} shows a qualitative schematic of this experiment. The results for a single test can be characterized using a confusion matrix (e.g. Figure \ref{fig:Ph_Ph}, description in Section \ref{sec:Ph_Ph}). The full $6\times6$ matrix of confusion matrices can be found in the Appendix \ref{sec:megamatrix} where the reader can zoom in on every individual test. In this section, we focus on the matrices corresponding to specific tests from the main handshake which address our core questions.

\subsubsection{Case study: training and testing with ideal \textsc{Photometry}}\label{sec:Ph_Ph}

\begin{figure}
	\includegraphics[width=\linewidth]{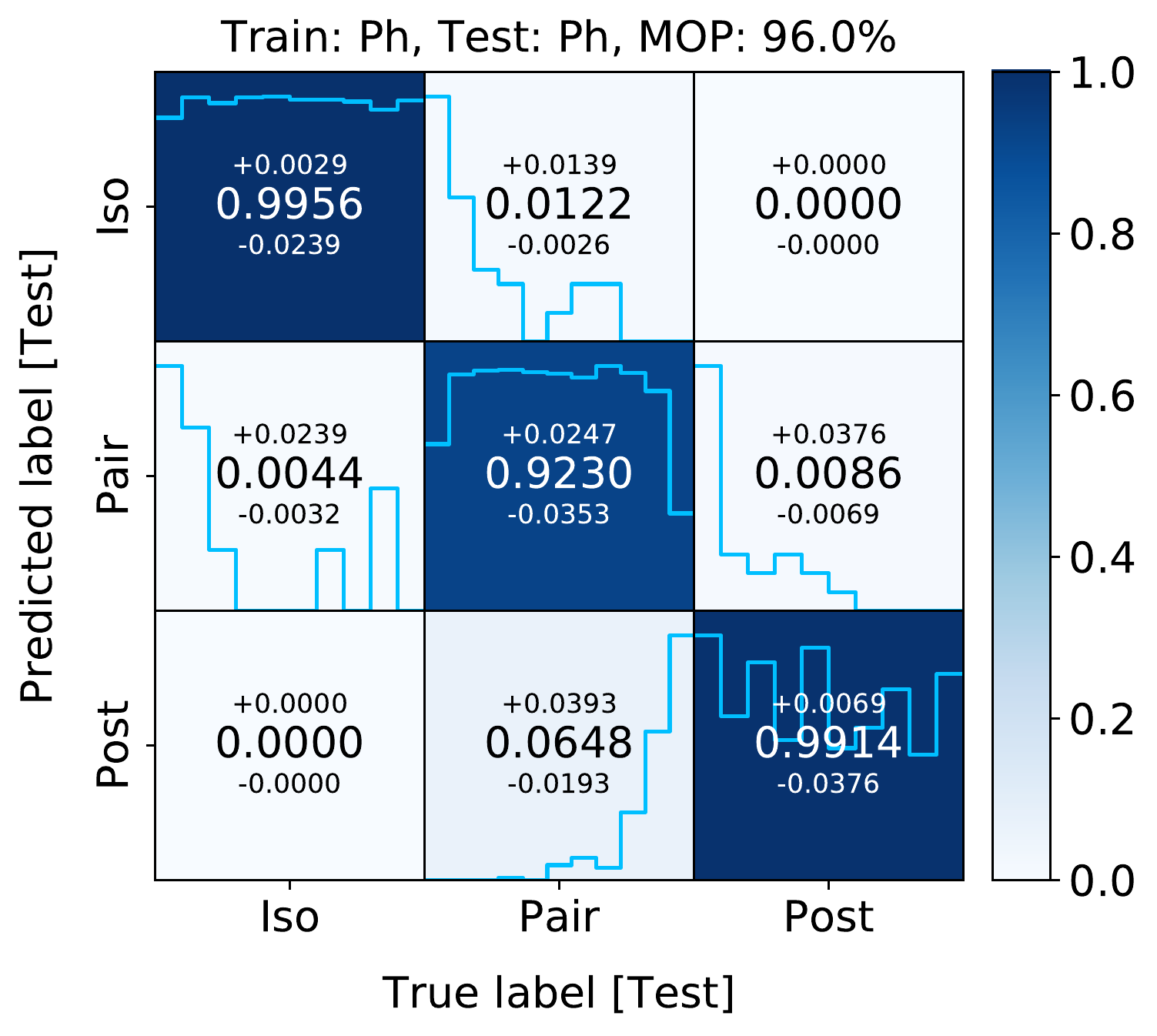}
    \caption{Confusion matrix for merger stage classifications using the \textsc{Ph} networks and corresponding test images. The number in each matrix element quantifies the median predicted fraction of images and 16$^{\mathrm{th}}$ and 84$^{\mathrm{th}}$ percentile offsets computed by bootstrapping the results of 10 unique realizations of training, validation and test data. Inset in each matrix element is a light blue bar plot showing the median empirical distribution function of relative merger class timescales (see Equation \ref{eq:timing}) for the images which fall in that matrix element in each of the 10 bootstraps. These timing histograms in each element are normalized by their respective maximum values for visibility in cases where the total number of images in a particular matrix element is small. Networks trained on \textsc{Ph} training data have a median overall performance (MOP) of $96.0\%$ when applied to \textsc{Ph} test data.}
    \label{fig:Ph_Ph}
\end{figure}

Figure \ref{fig:Ph_Ph} shows a single confusion matrix corresponding to the models trained on \textsc{Ph} and tested on \textsc{Ph} (as described in Section \ref{sec:Ph} and summarized in row 4 of Table \ref{tab:types}). We use this as a simple case study to orient the reader to the general features of our analysis for subsequent experiments. Each column of the confusion matrix shows the normalized distribution of predicted labels ($y$-axis) for all test images with a particular truth label ($x$-axis). A perfect classification network would therefore produce an identity matrix where all of the power is in the diagonal elements. Off-diagonal elements correspond to misclassifications denoted by a combination of predicted and true labels. Values shown in each matrix element are the median and 16th and 84th percentile range computed from the 10 bootstrap realizations of training, validation and test data. 

Figure \ref{fig:Ph_Ph} demonstrates that exceptional performance is achievable under ideal but generally unrealistic conditions: (1) noiseless images; (2) high spatial resolution; (3) no contamination by projection effects or other objects in the field of view; and lastly (4) a training set that is (by construction) generalizable to the test images. The ideal \textsc{Ph} networks have a median overall performance (MOP) of 96.0\%. Nonetheless, $7.7\%$ of pairs are misclassified by the networks -- with $6.5\%$ being misclassified as post-mergers and $1.2\%$ misclassified as isolated galaxies. Similarly, a small number of isolated and post-merger images are misclassified as pairs.\footnote{However, a subtle but notable feature of the networks is that the 55" fields of view (50 kpc at z=0.046) of the images often do not contain both galaxies for pairs. Nonetheless, networks distinguish pairs from isolated and post-merger images with great accuracy in these cases.}

Our confusion matrix includes an additional dimension which allows us to more deeply investigate network systematics and misclassifications. Histograms are inset in each matrix element which show the distributions of \emph{relative merger class timescales}:
\begin{align}\label{eq:timing}
t_{\mathrm{rel,iso}}  &= \frac{t - t_1 }{t_{10} - t_1}, \quad t_{\mathrm{rel,pair}} = \frac{t - t_{p}}{t_{c} - t_{p}}, \quad t_{\mathrm{rel,post}} = \frac{t - t_{c}} {t_{\mathrm{last}} - t_c}
\end{align}  
Here, $t$ is the simulation time-stamp associated with a particular snapshot in any simulation run. For isolated runs, $t_1$ and $t_{10}$ are the timestamps for first and tenth of the 10 snapshots selected. For merger simulations, $t_p$ is the time of first pericentric passage, $t_c$ is the coalescence time, and $t_{\mathrm{last}}$ is the timestamp of the last snapshot selected for a given run. $t_{\mathrm{last}}$ is therefore a number between $t_c+[250,500]$ Myr as per our class definitions and merger selection criteria. These normalizations allow us to place the timestamp of each snapshot (from each simulation) on a \emph{relative} timeline corresponding to its target class. An image from the isolated or post-merger classes has a $t_{\mathrm{rel,iso}}$ or $t_{\mathrm{rel,post}}$, respectively, that is between 0 and 1 by definition -- which we divide into 10 bins each. An image from the pair phase has $t_{\mathrm{rel,pair}}$ between -0.1 and 1 because (a) we start the pair phase 100 Myr before first pericentre and (b) the shortest $t_c-t_p$ is roughly 1 Gyr. Accordingly, each pair timing histogram has 11 bins and starts at 100 Myr before first pericentre and ends at coalescence. With these definitions, a uniform timing distribution in any matrix element would indicate that there is no temporal preference for the images assigned to that element. For visibility, the timing histograms in each element are normalized by their maximum values rather than the total number of images with the corresponding truth label.

Despite the good overall performance of this network and small fraction of misclassifications, the timing histograms reveal temporal preferences for misclassifying certain classes. True pairs that are misclassified as isolated are predominantly in the very early pair phase -- with the largest fraction in the pre- first pericentre bin (middle-top panel of Figure \ref{fig:Ph_Ph}). In contrast, true pairs that are misclassified as post-mergers are preferentially near coalescence (middle-bottom panel of Figure \ref{fig:Ph_Ph}). Consequently, the distribution of $t_{\mathrm{rel,pair}}$ for the correctly classified pairs is truncated in the first and final bins. The choppy $t_{\mathrm{rel,post}}$ distribution for the correctly classified post-mergers is due to the chance temporal resonance of snapshots selected for each merger. Coarser binning reveals an essentially uniform timing distribution. Lastly, post-mergers that are misclassified as pairs show a strong preference towards snapshots shortly after coalescence (right-middle panel of Figure \ref{fig:Ph_Ph}). 

The timing histograms in Figure \ref{fig:Ph_Ph} show that images which correspond to snapshots at the temporal interface between two neighbouring classes are the most challenging for the network to accurately classify. The subtlety is that these misclassifications are not completely spurious but rather follow intuitive temporal distribution functions -- which is actually a validation that the network is behaving as it should. For example, the most common mis-classification for pairs early in their interaction is "isolated". Likewise, the most common misclassification of late stage pairs, shortly before coalescence, is "post-merger". Conversely, it is rare for early pairs to be misclassified as post-mergers, or for late stage pairs to be classified as isolated. The misclassifications arise because the features of images on either side of a particular class boundary are genuinely similar. Indeed, the timing distributions for correctly and incorrectly classified pair and post-merger targets are qualitatively similar in our other tests (except where additional systematics due to strongly contrasting network/data types dominate). 

However, the timing distribution of isolated galaxies that are misclassified as pairs is less intuitive. First, the timing distribution of \emph{correctly} classified isolated galaxies is largely uniform. This result is important. Temporarily discounting secular changes to morphology (such as the emergence of bars and spiral arms), the main changes to these galaxies are their star-formation rates (SFRs) -- which decay exponentially with time. The uniform timing distribution for correctly classified isolated galaxies indicates that \emph{most} isolated galaxies are being correctly classified despite significant changes in SFR. However, the timing distribution of isolated galaxies that are misclassified as pairs suggests that early (and incidentally high-SFR) snapshots from the isolated runs are favoured. 

\begin{figure*}
	\includegraphics[width=0.495\linewidth]{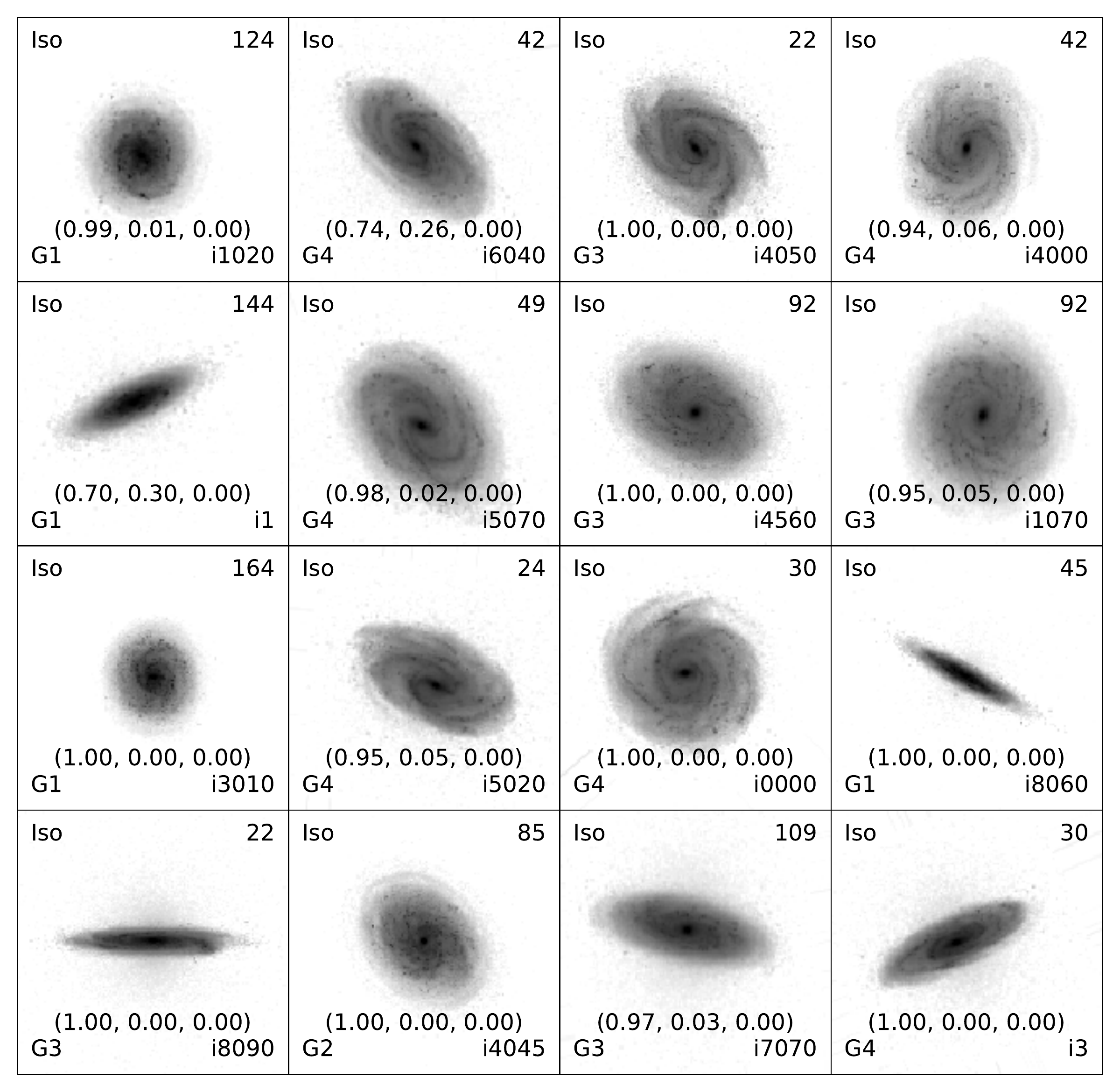}
	\includegraphics[width=0.495\linewidth]{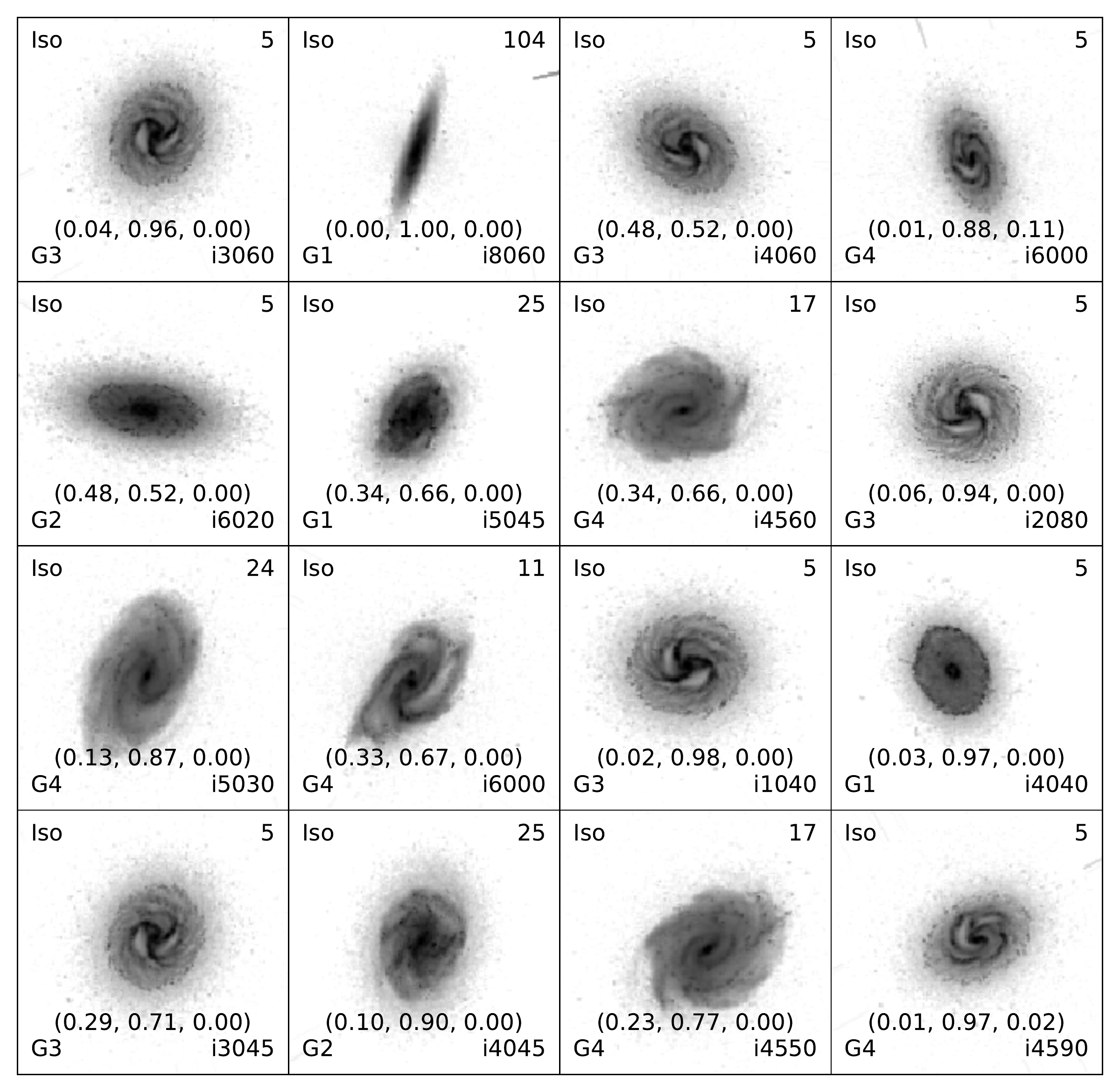}
    \caption{Randomly selected $r-$band idealized \textsc{Ph} images of isolated galaxies: (left) correctly classified as isolated; (right) incorrectly classified as pairs by the networks from Figure \ref{fig:Ph_Ph}. The unique identifier for each image is composed of the four labels in the corners of each image. The upper left shows the simulation type (all Iso in this case), the upper right is the snapshot number, the lower left is the galaxy ID, and the bottom right is the camera angle ID. The normalized probability distribution for each image is given as the tuple of ($P_{\mathrm{Iso}}$, $P_{\mathrm{Pair}}$, $P_{\mathrm{Post}}$). These single-band images show that there are notable differences in the surface-brightness distributions between correctly classified isolated galaxies and those misclassified as pairs. Many of the misclassified isolated galaxies have not yet relaxed from the initial conditions of the Iso simulation runs and exhibit non- steady-state morphologies. Streaks at the edges of some images are artifacts of border-handling when images are rotated/shifted during augmentation.}
    \label{fig:Iso}
\end{figure*}

Figure \ref{fig:Iso} shows 16 randomly selected images of isolated galaxies that are correctly classified as isolated (left panel) and incorrectly classified as pairs (right panel). The comparison reveals that, occasionally, misclassified isolated galaxies are not easily visually distinguished from the correctly classified isolated targets (e.g. first row, second column of the right panel). However, the right panel of Figure \ref{fig:Iso} shows that the majority of misclassified isolated galaxies have not yet dynamically relaxed from the initial conditions of the simulations and consequently have non- steady-state morphologies. Many have unusually bright spiral arms or rings of star-formation which may confuse a network which would desirably exploit morphological features such as tidal tails and shells to identify pairs. In addition, galaxies in the early stages of the merger simulations are similarly unrelaxed and should increase confusion between the pair and isolated classes. Again, there is a subtle importance to these results. The mischaracterization of these dynamically unrelaxed isolated galaxies as pairs \emph{confirms} that the network is exploiting desirable morphological features to make pair classifications. One may also visually note that (unlike morphology) high central surface brightnesses (such as those induced by a starburst) do not necessarily translate to a high pair probability. In Section \ref{sec:NC1}, we perform additional tests outside the main handshake which allow us to examine the temporal misclassification of isolated galaxies more deeply.

\subsubsection{Is radiative transfer necessary?}\label{sec:SMvsPh}

\begin{figure*}
\begin{center}
	\includegraphics[width=0.53\linewidth]{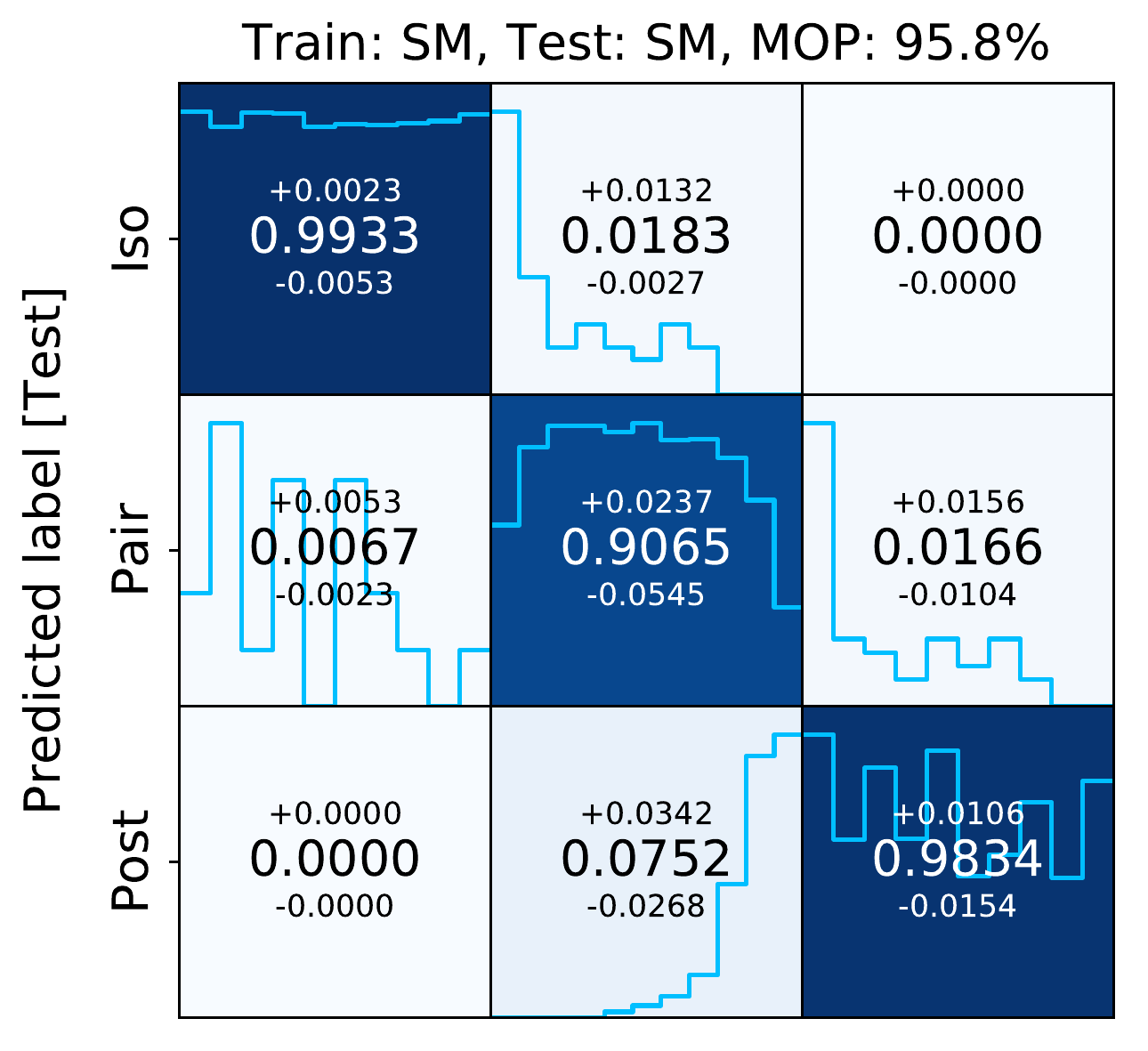}
	\includegraphics[width=0.4625\linewidth]{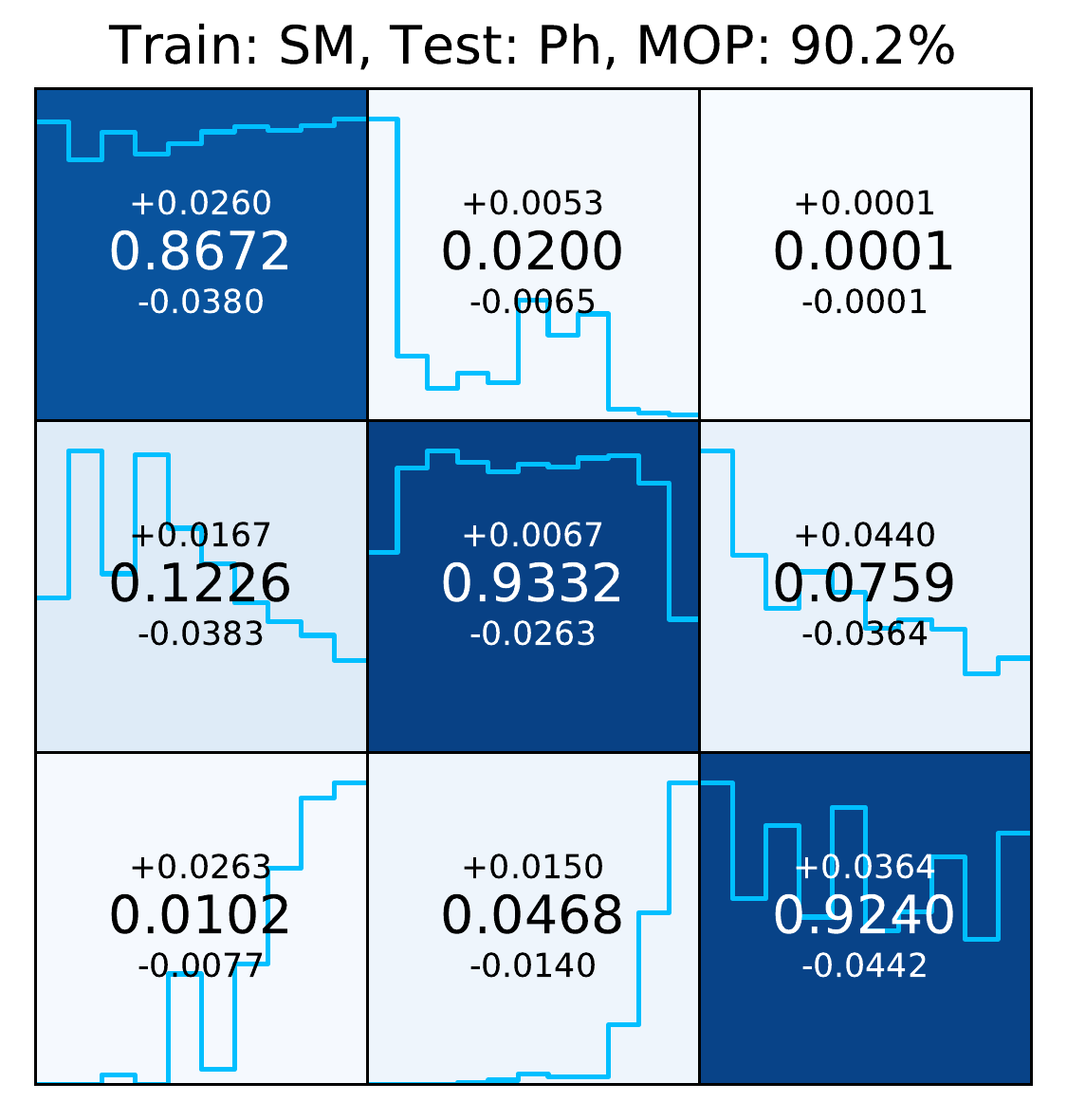}\\ 
	\vspace{-0.30cm}
	\includegraphics[width=0.53\linewidth]{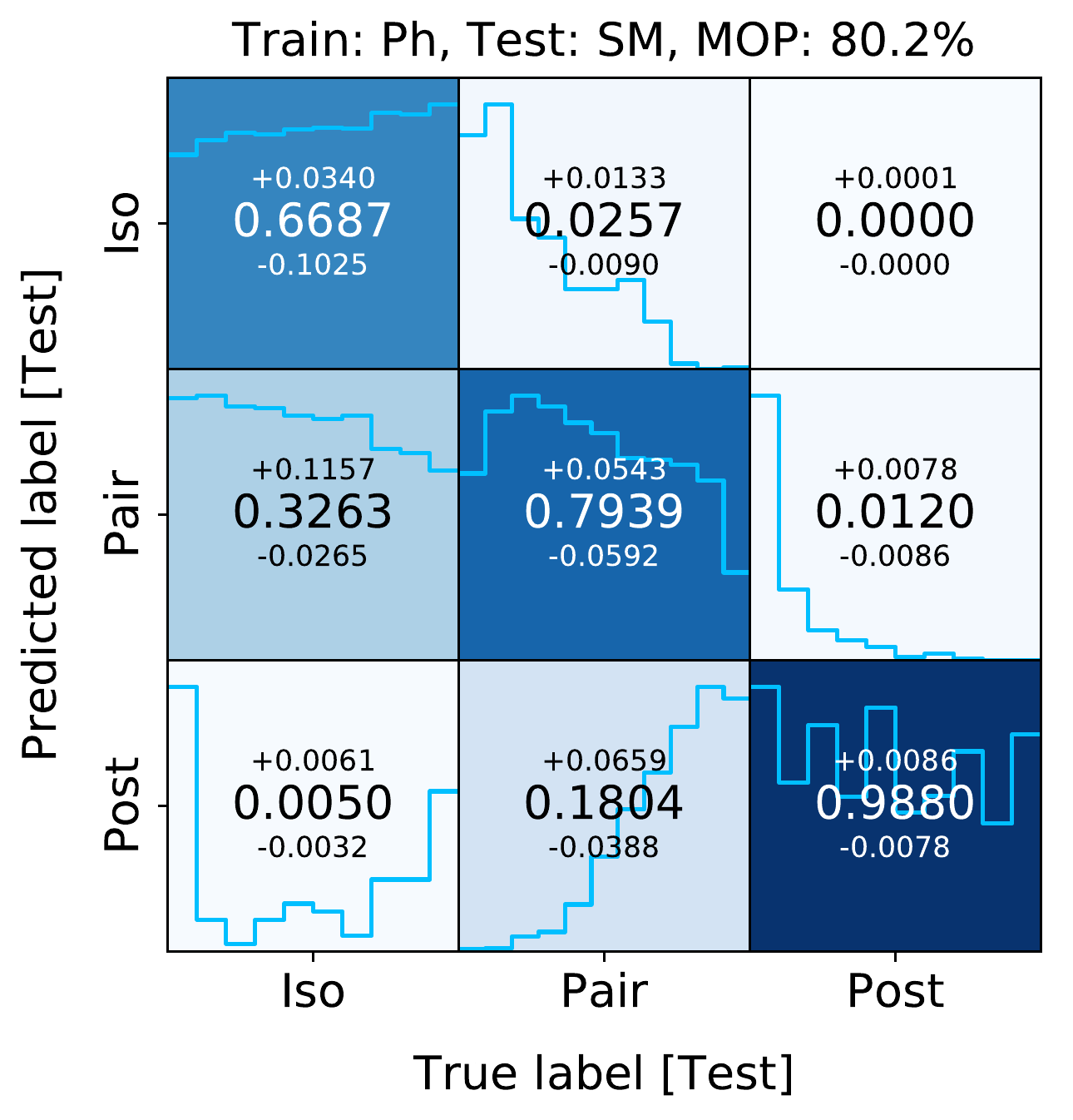}
	\includegraphics[width=0.4625\linewidth]{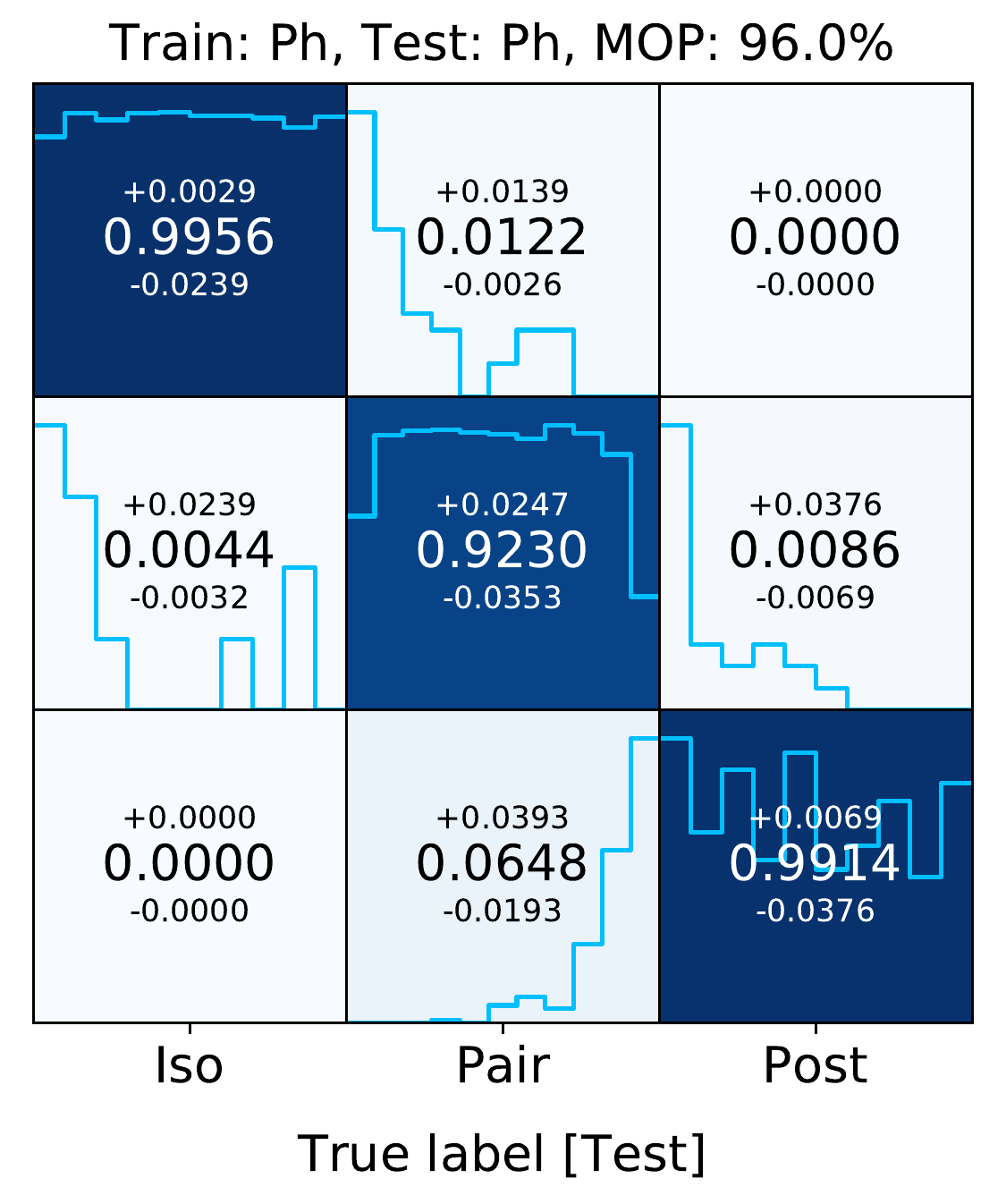}
    \caption{The importance of radiative transfer. Confusion matrices for merger stage classifications with \textsc{Sm} and \textsc{Ph} images/models. Models for matrices in the top row are trained on \textsc{Sm} images and matrices in the bottom row are trained on \textsc{Ph}. Models for matrices in the left column are tested on \textsc{Sm} images. Models for matrices in the left column are tested on \textsc{Ph} images. The columns of each matrix show the normalized distribution of predicted labels for each true label. The median and 16$^{\mathrm{th}}$ and 84$^{\mathrm{th}}$ percentile offsets are computed by bootstrapping the results of 10 unique realizations of training, validation and test data.}
    \label{fig:SM_Ph}
\end{center}
\end{figure*}

We are particularly interested in knowing whether \textsc{Sm} images are an adequate replacement for \textsc{Ph}. Radiative transfer makes photometric images computationally expensive to produce in large quantities and the data products from radiative transfer can be very large depending on the desired spectral resolution. For reference, in our dataset a single datacube with $(512\times512)$ spatial elements, a modest 241 spectral elements and 32-bit floating precision is 252.7 MB. Constructing a sufficiently large training set for application to cosmological volumes ($\sim10^6$ images) with these standards is a data management expense on the order of hundreds of Terabytes for the raw data products alone. Therefore, if a neural network performs equally well when trained on \textsc{Sm} images (1.05 MB/image for the same spatial resolution) as with images produced using radiative transfer (i.e. generation of \textsc{Ph}), it would not only save a significant step in the image generation pipeline, but also be a massive computational saving.

Figure \ref{fig:SM_Ph} shows the confusion matrices corresponding to our test of the importance (or not) of including radiative transfer and the generation of multi-band photometric images. The lower right panel shows the same high-performance result as in Figure \ref{fig:Ph_Ph} where the idealized \textsc{Ph}-trained networks are applied to \textsc{Ph} test data. The upper left panel shows that nearly equally high performance is achieved by the \textsc{Sm} networks on \textsc{Sm} test data. The off-diagonal panels show the results of testing of these networks on data from the other type. The lower left panel shows that networks trained on \textsc{Ph} and tested on \textsc{Sm} images have significantly lower performance (median performance of $80.2\%$) than when either \textsc{Sm} or \textsc{Ph} networks are tested on data of their own respective type. In contrast, the upper right panel shows that networks that are trained on \textsc{Sm} images and tested on \textsc{Ph} still have excellent performance (median performance of $90.2\%$). These results are intuitive when we reflect on the differences between a single-band photometric image and a map of stellar mass. With respect to the \textsc{Sm} images, \textsc{Ph} images include higher-order information from which the network can draw (such as locally varying mass-to-light ratios due to the ages and metallicities of stellar populations and dust). If these higher-order features correlate with the target classifications then the \textsc{Ph} network may suffer from an unconventional form of overfitting with respect to the corresponding \textsc{Sm} images -- because these higher-order features are absent in the \textsc{Sm} images. 

In contrast, the morphological disturbances that are exploited by the \textsc{Sm} network when training on \textsc{Sm} images will always be present in the \textsc{Ph} images. They will simply underlay any higher-order \textsc{Ph} features. Consequently, the \textsc{Sm} model tests with higher performance on \textsc{Ph} images than vice-versa because the \textsc{Sm} network is \emph{guaranteed} to focus on lower-order features and thus is generalizable to \textsc{Ph}. Later, in Section \ref{sec:limitations}, we argue that the disparity between \textsc{Ph} and \textsc{Sm} networks/data arises due to the limitations of our training data and predict it would disappear with a galaxy population that is more diverse in stellar populations, colours and gas fractions.

The results of this section demonstrate that radiative transfer provides a network with more exploitable features than are available in \textsc{Sm} images. These include higher-order features of the surface brightness profiles and the colour information that can be made accessible by producing images in multiple bandpasses. However, we also show that idealized \textsc{Sm} networks are, nonetheless, highly effective at handling idealized \text{Ph} images. This means that, at least in the idealized case, one can avoid the (potentially enormous) computational and data management expenses of radiative transfer for large datasets by using \textsc{Sm}-based images for a modest trade-off in performance. 

\subsubsection{Is observational realism necessary?}\label{sec:realism}

We are ultimately only interested in networks that will perform well on realistic images. In the last two sections, we have shown that the networks trained on idealized \textsc{Sm} and \textsc{Ph} images perform very well on their respective selves and reasonably well on each other. In this section, we address the question of whether networks trained with idealized images can accurately classify images with realistic noise, resolution degradation and contamination by nearby objects in the images' fields of view. Our benchmark for assessing how well any of our networks will perform on real data is the \text{Photometry FullReal} dataset -- which is our best representation of real data. The tests in this section are designed to tell us whether it is sufficient to construct idealized \textsc{StellarMap} or \textsc{Photometry} synthetic images as training data for networks that can be applied to real data. 

\begin{figure*}
	\includegraphics[width=0.355\linewidth]{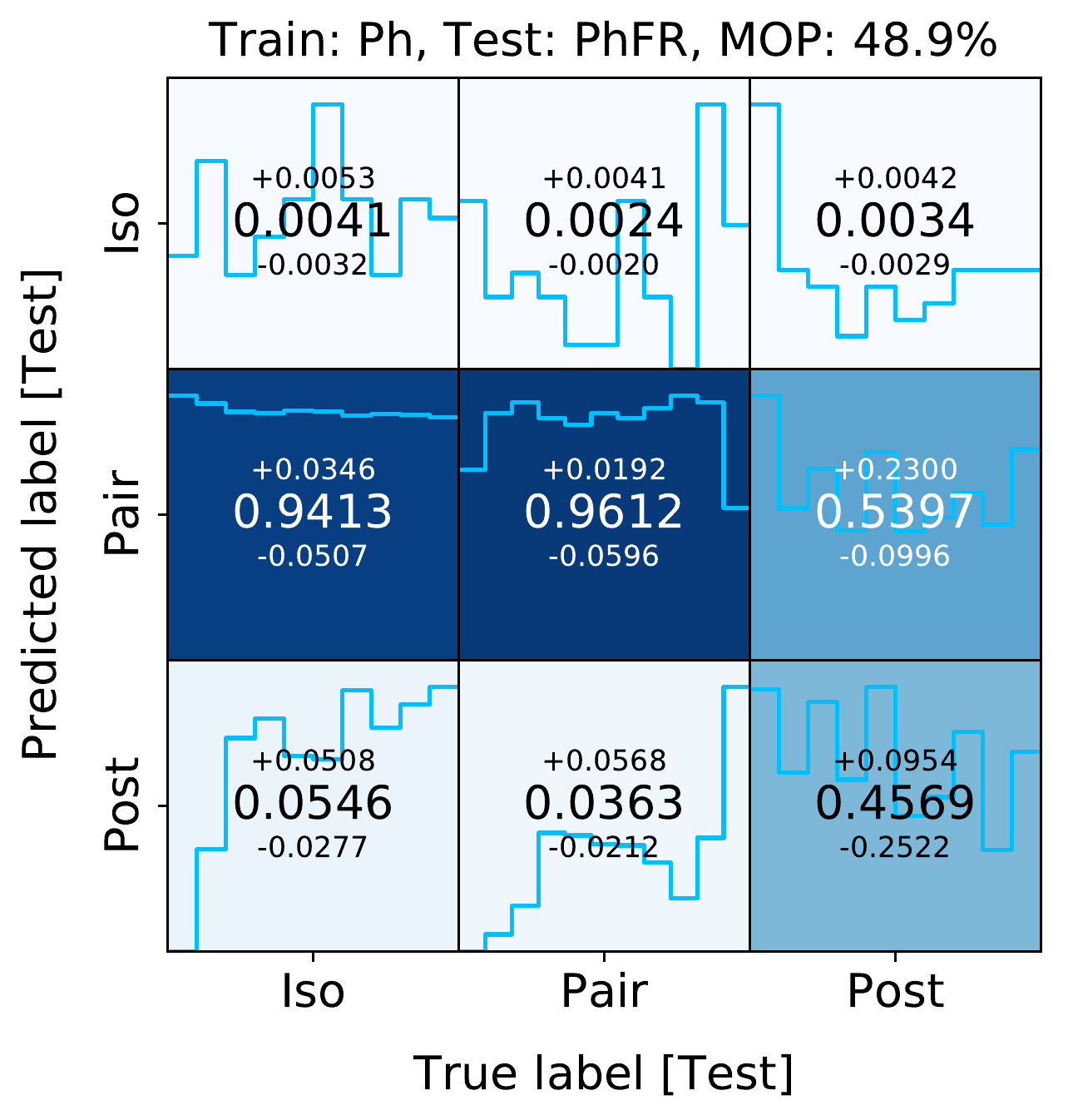}
	\includegraphics[width=0.31\linewidth]{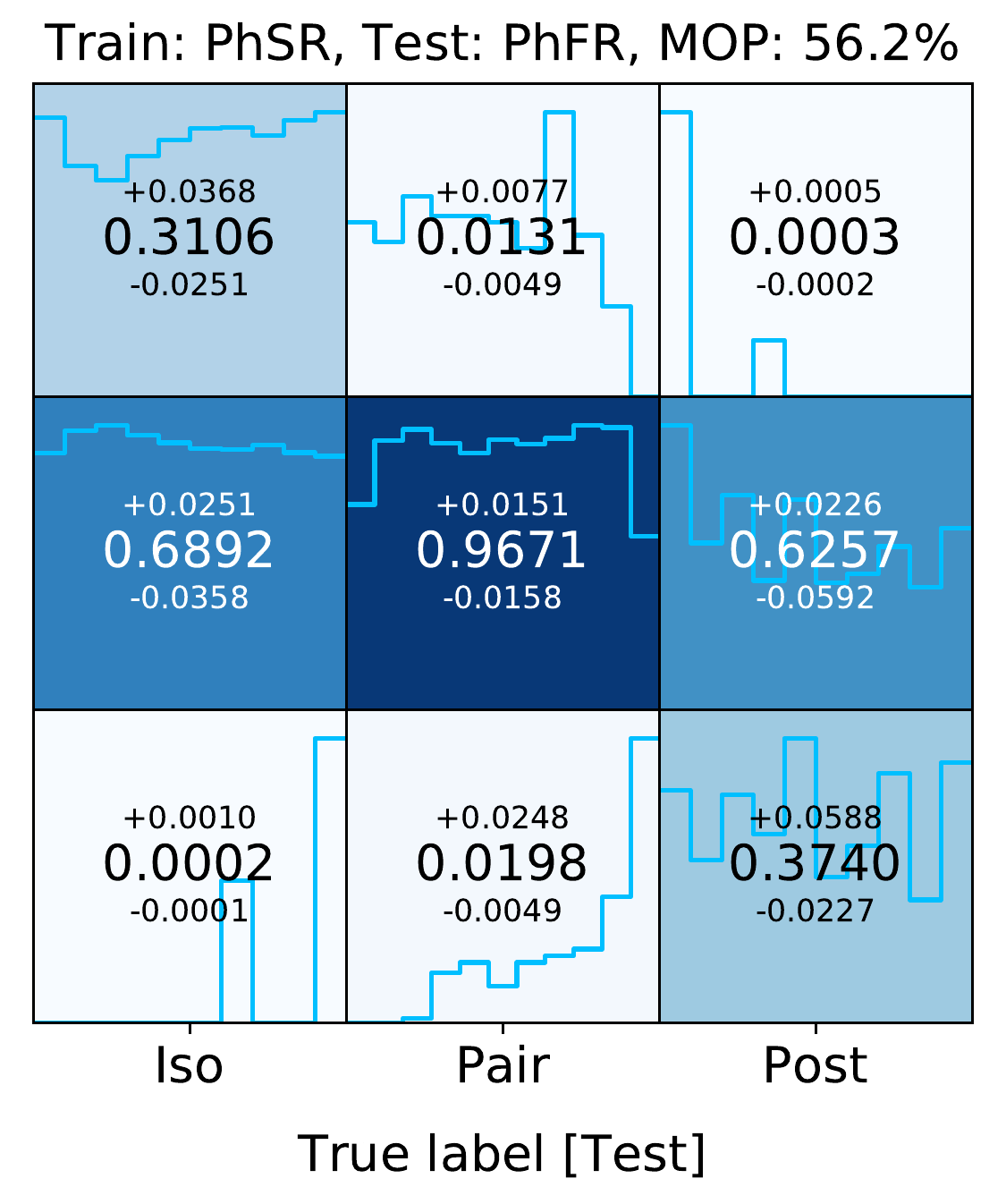}
	\includegraphics[width=0.31\linewidth]{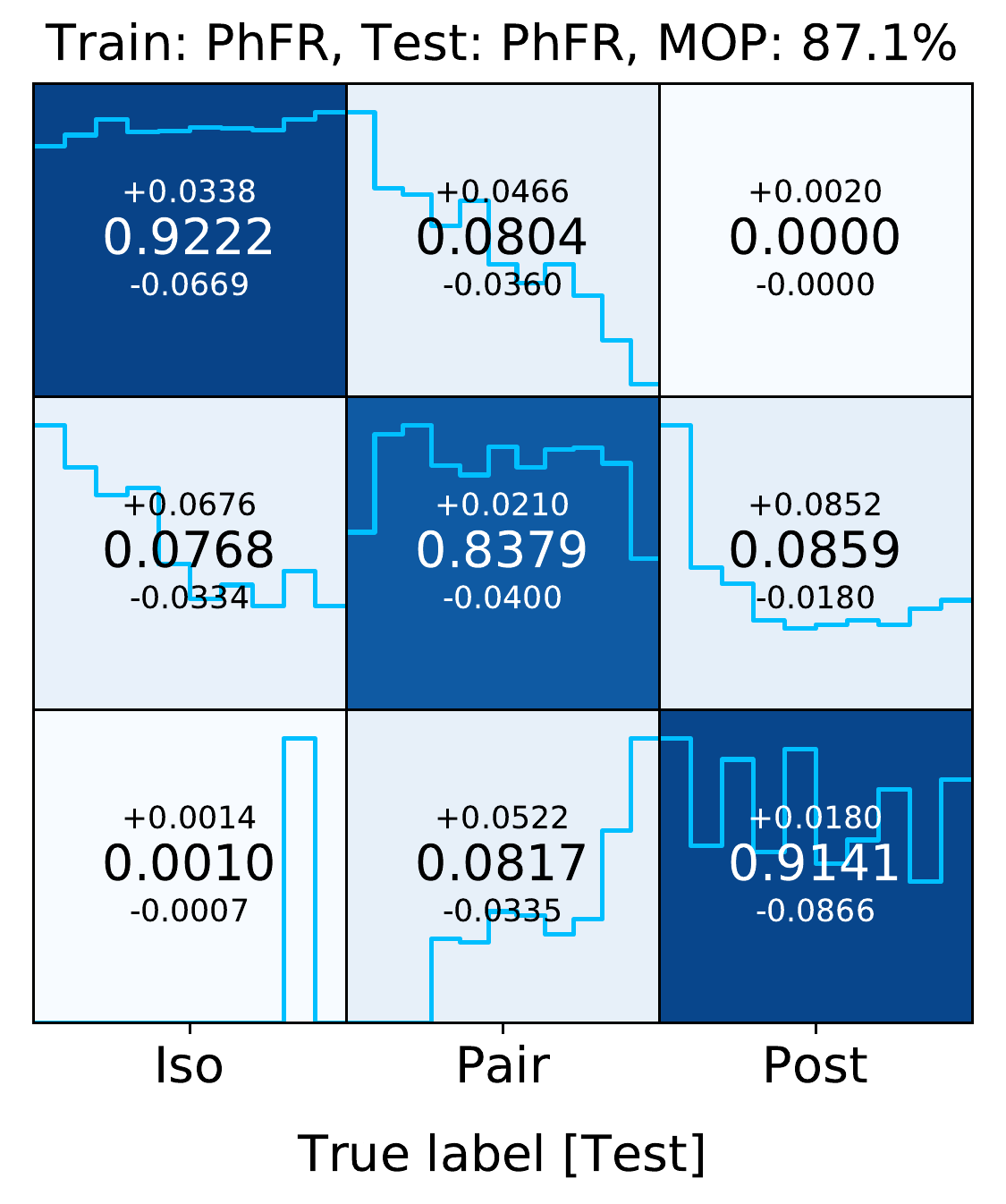}
    \caption{The importance of realism. \textsc{Ph} (left), \textsc{PhSR} (centre) and \textsc{PhFR} (right) networks are applied to the \textsc{PhFR} test data. Both the idealized \textsc{Ph} and \textsc{PhSR} networks systematically misclassify non-pair targets as pairs. In contrast, networks that are exposed to training data with full realism (real noise, resolution \emph{and} crowding) do not appear to be affected by this systematic and an overall median performance of $87.1\%$ is achieved with the \textsc{PhFR} networks on \textsc{PhFR} test data. }
    \label{fig:realism}
\end{figure*}

Figure \ref{fig:realism} shows the results of applying the \textsc{Ph} (left), \textsc{PhSR} (centre) and \textsc{PhFR} (right) trained networks to \textsc{PhFR} test data. In this sub-section, we will focus on the left and right panels -- which demonstrate the importance of realism, returning to the central panel of Figure \ref{fig:realism} in the following sub-section. The left panel shows that \textsc{Ph} networks have very poor performance when tested on realistic images. Similarly poor results are obtained using the idealized \textsc{Sm} networks (see corresponding panel in Figure \ref{fig:megamatrix} in Appendix \ref{sec:megamatrix}). Both idealized \textsc{Ph} and \textsc{Sm} networks systematically classify targets in the \textsc{FullReal} images as pairs. The histograms inset in the elements of each matrix show that there are no particular temporal preferences for post-merger or isolated targets that are misclassified as pairs by these networks.  

On the other hand, the right panel of Figure \ref{fig:realism} shows that networks that are trained using \textsc{PhFR} images perform very well on other \textsc{PhFR} survey-realistic images -- despite the full statistical rigour of noise, resolution and contamination effects. Indeed, these results suggest that it is \emph{only} because this network was exposed to these biases in training that it is capable of handling other realistic images. We investigate this hypothesis more closely in the next section. 

\subsubsection{Is the \emph{level} of realism important?}\label{sec:level}

In the previous sub-section, we showed that networks that are trained on either \textsc{Ph} or \textsc{Sm} idealized images perform poorly when tested on realistic images, whereas the \textsc{PhFR} network performed very well. To interpret these results, we now focus on the following question: what ingredients of the FullReal images are key to the success of the FullReal network? Are realistic skies and resolution sufficient criteria? Or are realistic additional sources necessary too? We answer these questions using the networks trained on SemiReal images which are tested on FullReal images. Recall that the sky noise and spatial resolution in the SemiReal images are statistically the same as for the FullReal images by construction (see Section \ref{sec:PhSR}). Consequently, the only difference between these two datasets is that the FullReal images can contain other objects in the field of view which may confuse a network that is not used to seeing additional sources.

The central panel of Figure \ref{fig:realism} shows the results of applying \textsc{PhSR} networks to \textsc{PhFR} test sets. Again, the results are qualitatively similar when we apply the \textsc{SmSR} networks to the \textsc{PhFR} or \textsc{SmFR} images (see corresponding panels in Figure \ref{fig:megamatrix} of Appendix \ref{sec:megamatrix}). These tests reveal that the \textsc{SemiReal} networks (whether originally deriving from \textsc{Sm} or \textsc{Ph} images) systematically classify targets as pairs in the \textsc{FullReal} images -- as was the case for the idealized \textsc{Ph} and \textsc{Sm} networks. The consequently poor overall performance, particularly when compared to the successes we see when training and testing using FullReal images (right panel of Figure \ref{fig:realism}), demonstrate that the \emph{level} of realism is crucial to network performance with realistic images. Specifically, without exposure to projection effects and additional sources of contamination during training, the \textsc{SemiReal} networks preferentially associate secondary sources in the images as companions to the target. In contrast, networks trained on images that include contaminating effects beyond sky and resolution degradation must learn ways to separate false positives and true positives with respect to the pair class. 

\begin{figure}
	\includegraphics[width=0.99\linewidth]{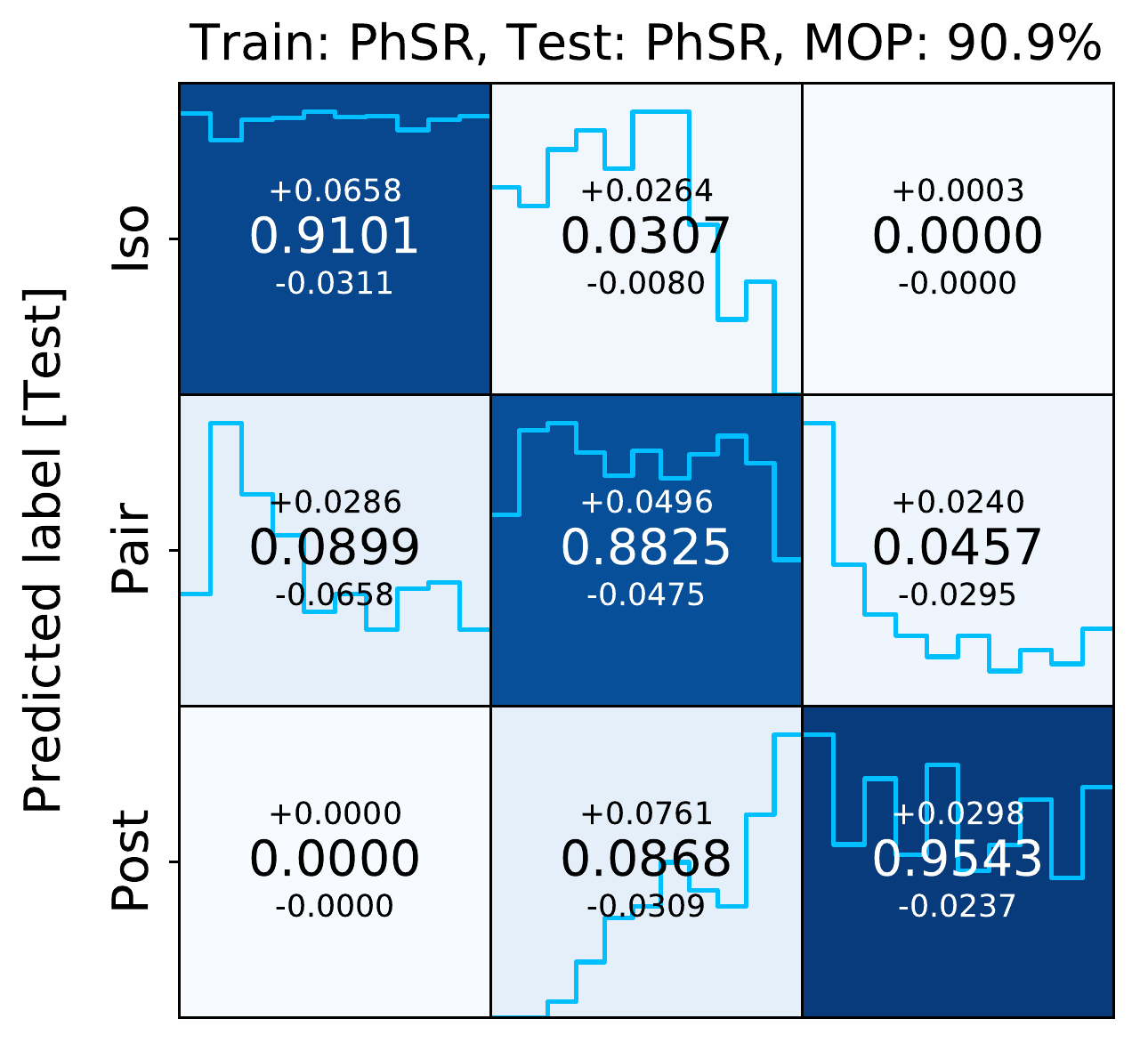}\\
	\includegraphics[width=0.99\linewidth]{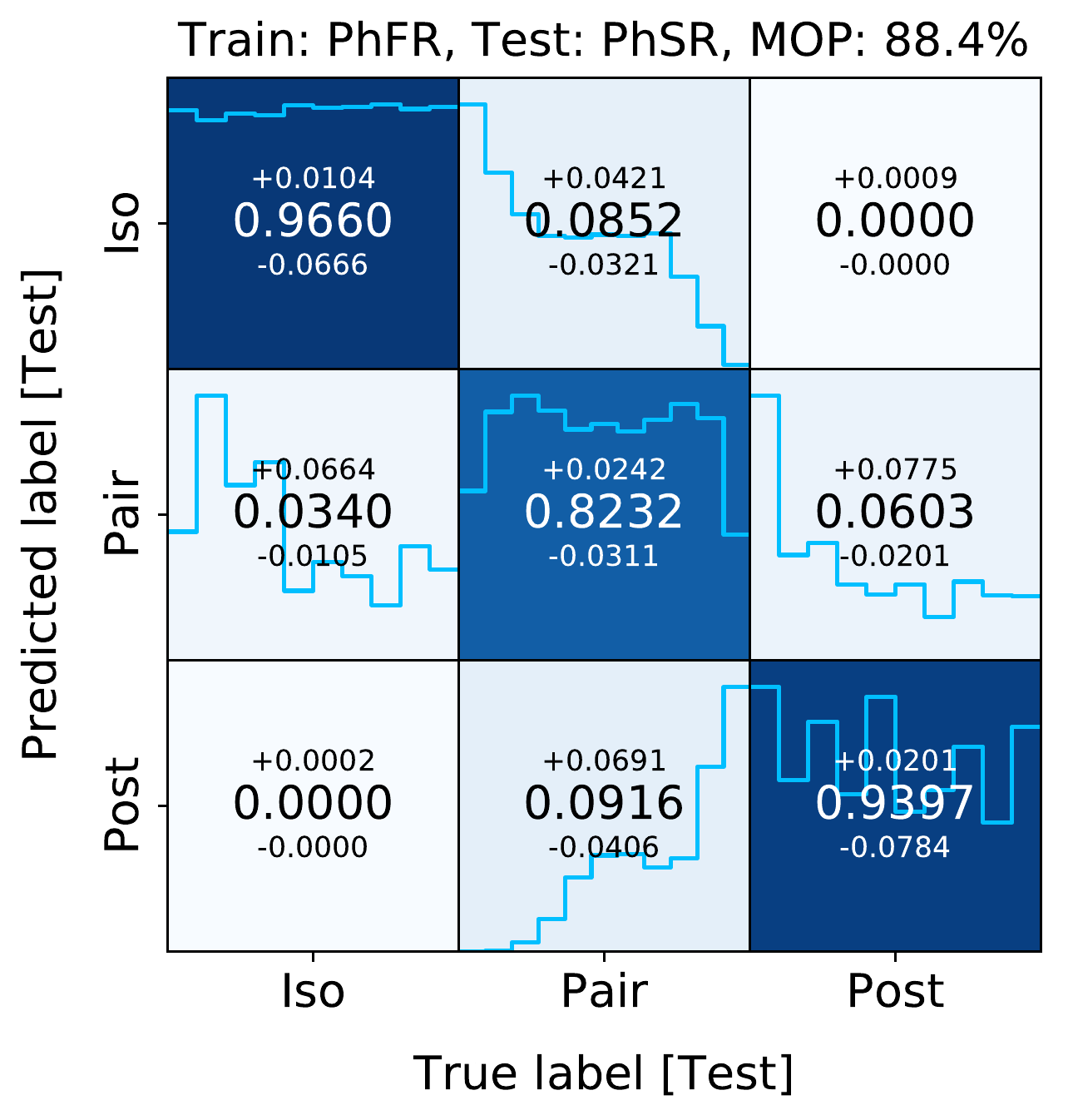}
    \caption{The importance of the \emph{level} of realism. The upper panel shows that \textsc{PhSR} networks (trained on images with realistic skies and resolution but no field objects) handle other \textsc{PhSR} images with very good performance. In contrast, the middle panel from Figure \ref{fig:realism} showed that \textsc{SemiReal} networks perform very poorly when applied to fully realistic images and systematically classify targets as pairs -- regardless of their true class. The lower panel here shows that the reverse of this test (training on \textsc{FullReal} images and testing on \textsc{SemiReal} images) produces even better results than when the \textsc{PhFR} networks to \textsc{PhFR} test data. This test shows that networks that are trained on images which include contaminated FOVs have no trouble handling images which have similar noise and resolution but do not contain contaminants.}
    \label{fig:level}
\end{figure}

Figure \ref{fig:level} examines some important details in the relationship between the level of realism in training data and network performance on realistic test images. The upper panel of Figure \ref{fig:level} shows that \textsc{SemiReal} networks perform very well on \textsc{SemiReal} test data -- while we know from the middle panel of Figure \ref{fig:realism} that this performance in uncontaminated FOVs does not translate to the \textsc{FullReal} images. However, the reverse of that test (training on \textsc{FullReal} images and testing on \textsc{SemiReal} images) would show whether the contaminants in the \textsc{FullReal} training data negatively affect network performance on images which do not contain contaminants. The lower panel of Figure \ref{fig:level} shows the results of applying \textsc{FullReal} networks to the \textsc{PhSR} test images. This test confirms that the networks trained on FullReal images have no trouble testing on images which have similar skies and resolution but which do not contain contaminants. Indeed, statistically fewer isolated ($-4\%$) and post-merger ($-2\%$) targets in the \textsc{SemiReal} test images are mis-classified as pairs by the \textsc{PhFR} network compared to the right panel of Figure \ref{fig:realism}. It is important to note that, until now, we have not seen a network that is trained on one image type and performs equally well (or better) on another image type. Given that our network architecture is the same for every model (with the exception of the number of input channels in particular cases), Figure \ref{fig:level} is also a crucial validation that the networks are not predestined to overfit to their own training data due to some property of the model architecture. 

The results of this section show that training images with realistic noise and resolution are important but (on their own) insufficient criteria for achieving high merger classification performance in realistic images. Networks will only learn to discriminate between true and false pairs if they have been exposed to realistic fields of view in their training data (\textsc{FullReal} images). As such, training data must include realistic noise and resolution \emph{as well as} contamination by additional sources. 

\subsubsection{Is realism more important than radiative transfer?}\label{sec:realorrad}

In the previous section, we showed that the highest performance on realistic test data is obtained when training with \textsc{PhFR} images (see Figure \ref{fig:realism}). In addition, in Section \ref{sec:SMvsPh} and Figure \ref{fig:SM_Ph} we showed that networks trained on \textsc{Sm} images performed well when tested on \textsc{Ph} data. So we return to the guiding question of Section \ref{sec:SMvsPh}: \emph{Can we get away with using \textsc{StellarMap}-based images in lieu of radiative transfer?} In other words, is the realism more important than whether the realistic image are originally derived from a \textsc{StellarMap} or from \textsc{Photometry}?

\begin{figure*}
	\includegraphics[width=\linewidth]{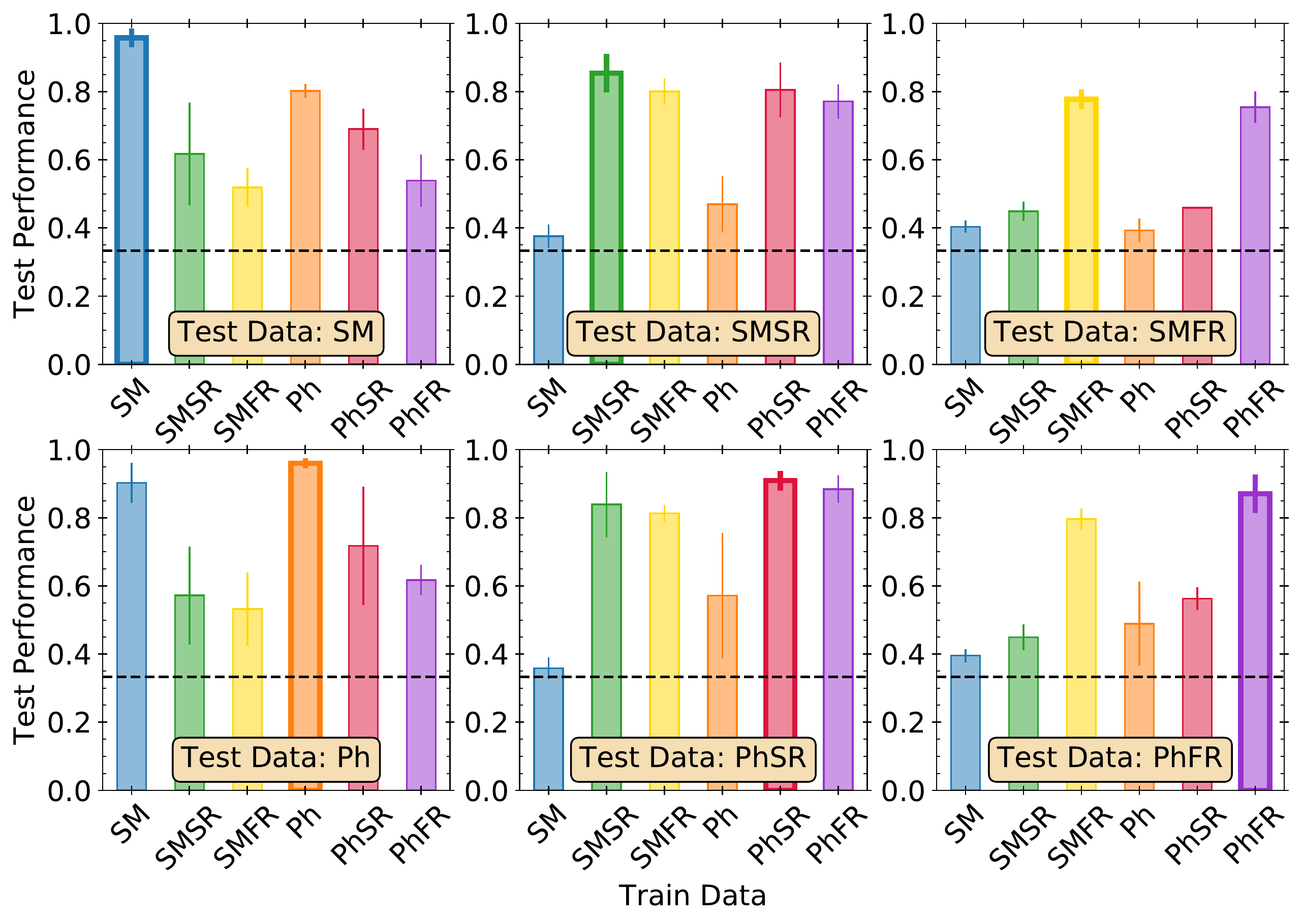}
	    \caption{Overall median test performances in our network/data handshake. Each panel shows the median test results of applying a specified network to test images of every type. The ordering of test data types are indicated along the $x$-axes. For reference, the overall performance of a single test is computed from the number of test images in the diagonal elements of a confusion matrix as a fraction of the total number of images. Median overall performance is computed over the 10 bootstraps of training/validation/test sets. Similarly, errorbars show the $5^{\mathrm{th}}$ to $95^{\mathrm{th}}$ percentile range in overall performance. Bolded borders are placed on bars corresponding to cases in which the test data is of the same image type as the training data. The dashed line denotes a uniformly random classification performance -- which in the case of three possible classes is $1/3$. The only model that performs comparably with the \textsc{PhFR} model on the \textsc{PhFR} test data is the \textsc{SmFR}. This result demonstrates that matching the realism is more important than whether or not the training data derives from images generated with proper radiative transfer.}
    \label{fig:testall}
\end{figure*}

Figure \ref{fig:testall} offers a compressed view of every test in the main handshake shown in Figure \ref{fig:megamatrix} of Appendix \ref{sec:megamatrix}. Figure \ref{fig:testall} shows the median overall performances of each network applied to each set of test data. The overall performance is computed as the number of images in the diagonal elements of a confusion matrix relative to the total number of images. Each panel shows the results of networks trained using each type of training data (labels along $x$-axis) and tested on a particular type of test data (indicated in the tan box). Coloured bars show the median and sample standard deviation test performance of each network type for the 10 networks trained using different random samplings of the training images. The dashed black line denotes a random performance of 1/3 for a three-class model. Since \textsc{PhFR} data are closest to what would be observed with a real instrument, the lower right panel is the focus of this section. As we showed in Figure \ref{fig:realism}, the \textsc{Sm} and \textsc{Ph} networks do only slightly better than random when applied to the \textsc{PhFR} test images because of the lack of realism. Similarly, we showed in Section \ref{sec:level} that the \textsc{SemiReal} networks do only marginally better than models with no realism because they are not exposed to projection effects in training. In contrast, \emph{both} \textsc{FullReal} networks (\textsc{SmFR} and \textsc{PhFR}) perform well on \textsc{PhFR} images. The \textsc{SmFR} and \textsc{PhFR} networks have median overall performances of $79.6\%$ and $87.1\%$, respectively, when applied to \textsc{PhFR} images. 

In contrast, networks trained on either idealized or \textsc{SemiReal} images never exceed $60\%$ performance when handling the more realistic PhFR images. This is true whether the training images are derived from photometry or stellar maps. These results show that the level of realism is more important than radiative transfer and that one can achieve strong performance with \textsc{Sm}-based images as long as they are fully realistic. Although there is a big difference between a network that can achieve $\sim90\%$ performance compared to one that achieves $\sim80\%$, the difference in performance may be an acceptable trade-off for being able to side-step radiative transfer and its associated computational and data management expenses -- particularly on the scale of the current state-of-the-art cosmological simulations.

\subsection{Single-channel experiments}\label{sec:NC1}

We supplement the main handshake experiment with a series of additional single-band tests with \textsc{Ph}-based networks. These tests are designed to determine the importance of colour and bandpass to network performance. In particular, we are interested in whether the timing preference for misclassified isolated galaxies seen in several \textsc{Ph}-based tests (e.g. Figure \ref{fig:Ph_Ph} and the right panel of Figure \ref{fig:realism}) and discussed in Section \ref{sec:Ph_Ph} persists for networks that are colourblind. If the timing preference persists and similar performance is achieved when a network is trained using a single band, then colour can be ruled out as a major factor in distinguishing pairs from isolated galaxies by the network. 

More generally, we are also interested in knowing the degree to which overall network performance is sensitive to colour. There are important advantages of a network that can achieve high performance without exploiting colour and focuses primarily on morphological features. For example, star-formation correlates strongly with colour. Since interactions between gas rich galaxies are proven triggers of central star formation \citep{1987ApJ...320...49B,1991MNRAS.251..360N,1994ApJ...435..540C,1994ApJ...431L...9M,1996ApJ...464..641M,Barton_2000,2000MNRAS.312..859S,2007AJ....133..791S,2008MNRAS.384..386C,2008AJ....135.1877E,2011MNRAS.412..591P,2013MNRAS.430.1901H,2013MNRAS.433L..59P,2015MNRAS.448.1107M,2016MNRAS.462.2418S,2019MNRAS.482L..55T}, a colour-sensitive network may learn to exploit central star-formation to characterize interaction stage through its correlation with colour. However, the negative consequence of identifying/characterizing interactions based on triggered star-formation is that any study which then examines the relationship between interaction stage and star-formation is automatically biased. Therefore, it is of great value to know that our networks are able to make merger-stage classifications without exploiting colour information.

The single-channel experiments are divided into two handshakes. New \textsc{Ph}, \textsc{PhSR} and \textsc{PhFR} networks are each trained and tested using only the corresponding $r$-band and $i$-band images to produce a single $3\times3$ handshake for each band. As with the main handshake, we statistically combine the individual test results of 10 bootstraps of training/validation/test images for our final results. The results of every test are shown in Appendix \ref{app:NC1}. We discuss selected results in the subsections that follow.

\begin{figure}
	\includegraphics[width=0.99\linewidth]{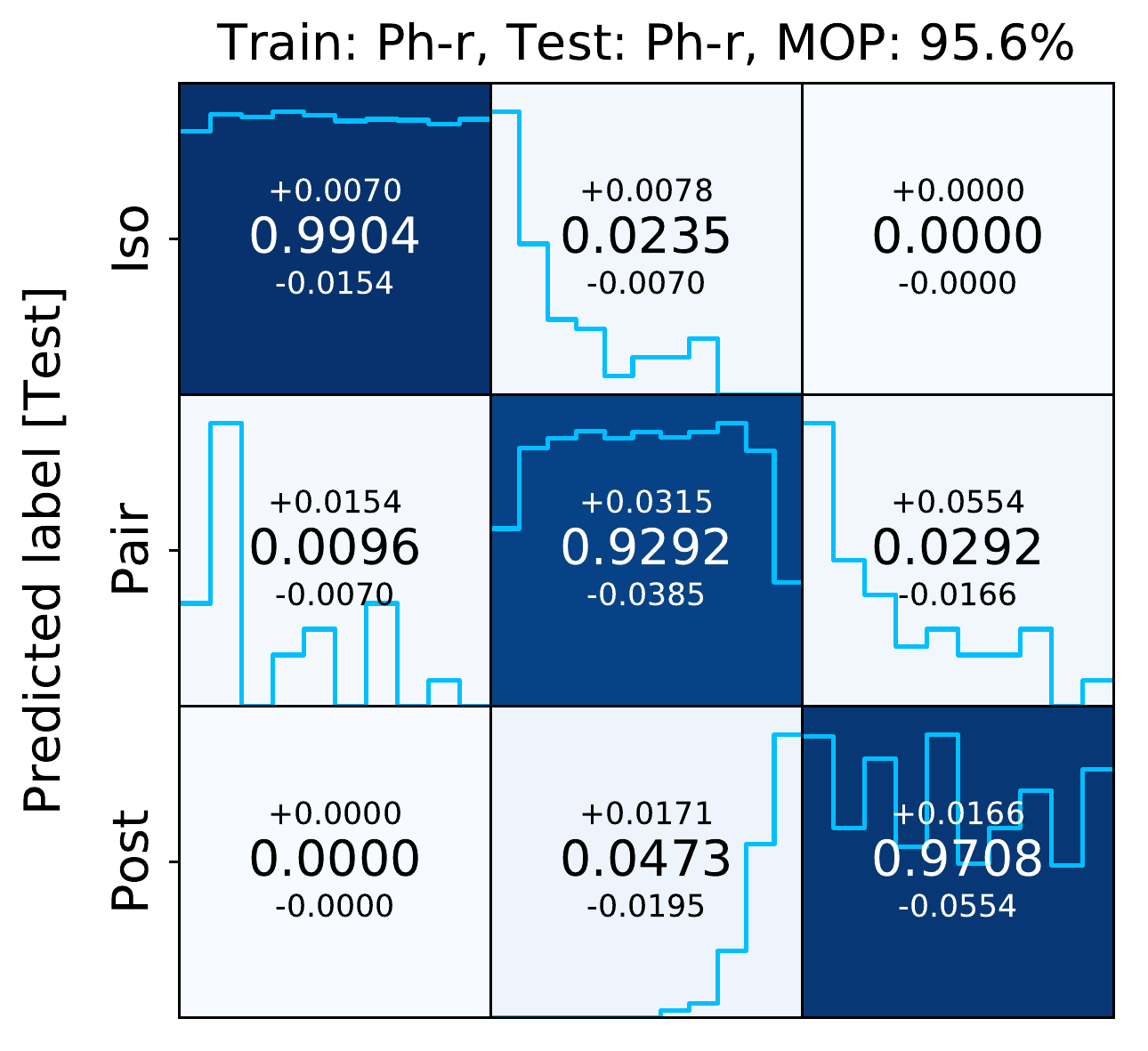}
	\includegraphics[width=0.99\linewidth]{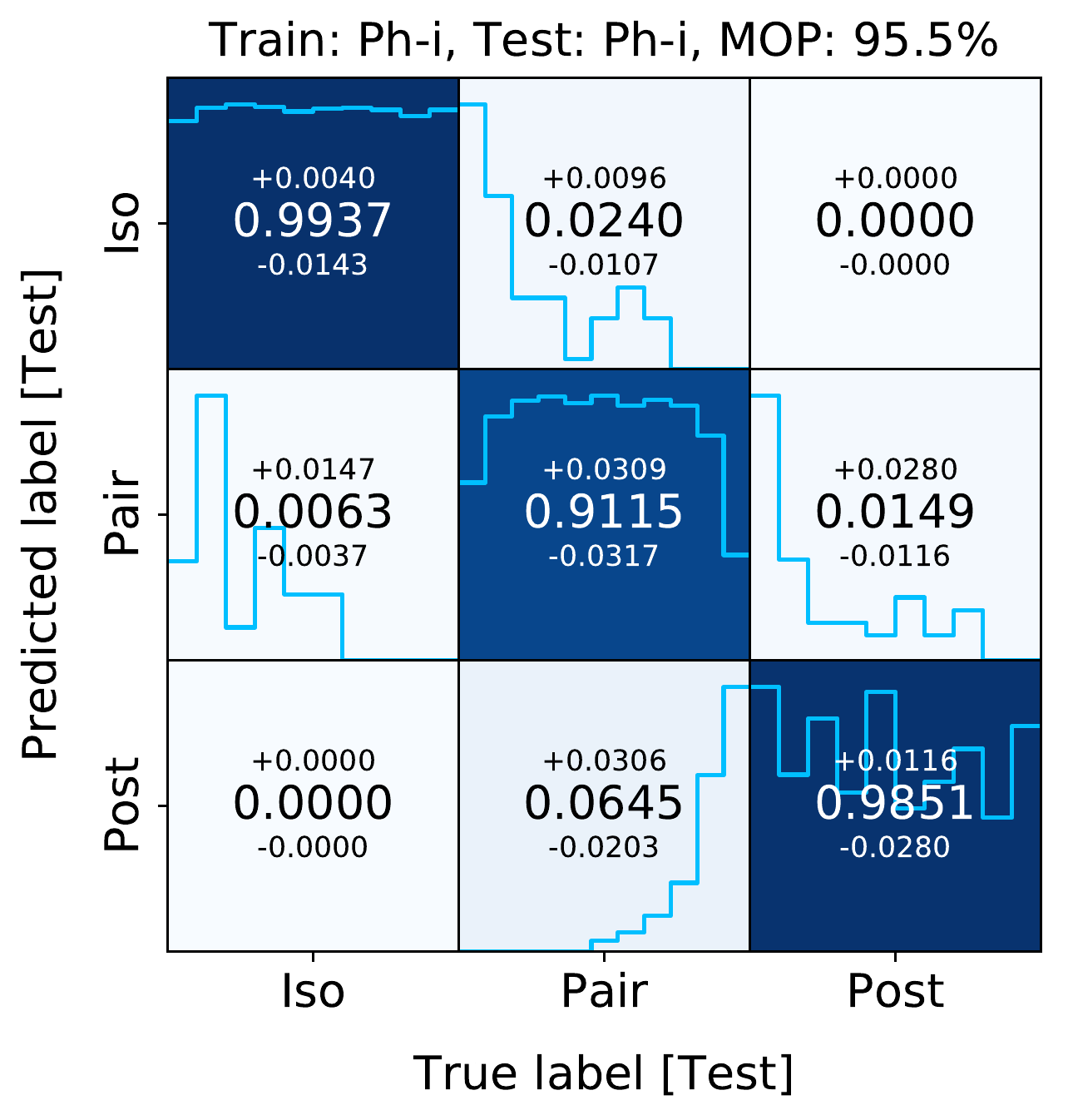}
    \caption{The importance of colour to network performance with idealized images. Network results shown in the upper and lower panels were trained and tested on single-channel $r$- and $i$-band idealized \textsc{Ph} images, respectively. The networks achieve median overall performances of $95.6\%$ ($r$-band) and $95.5\%$ ($i-$band) compared to $96.0\%$ when training and testing using idealized images from all three $gri$ bands. The losses in performance when reducing to single-band photometry are minor.}
    \label{fig:NC1}
\end{figure}

\subsubsection{How important is colour to network performance?}\label{sec:colour}

Previous tests already give us reason to believe that colour is not an \emph{essential} ingredient of satisfactory network success. For example, Figure \ref{fig:testall} showed that even the \textsc{SmFR} networks (which do not use radiative transfer) achieve a reasonable median overall performance of $79.6\%$ on \textsc{PhFR} test images using only single-channel input. Similarly, the upper right panel of Figure \ref{fig:SM_Ph} showed that the colourblind, idealized \textsc{Sm} network achieved $90.2\%$ median overall performance when tested on idealized \textsc{Ph} test images -- only a $5.8\%$ drop with respect to the results of the full-colour \textsc{Ph} network.

Figure \ref{fig:NC1} shows the confusion matrices for the idealized $r$-band (upper panel) and $i$-band (lower panel) \textsc{Ph} networks which were tested on the respective $r$-band and $i$-band \textsc{Ph} test images. For the idealized networks/data, the single-channel \textsc{Ph} networks still achieve exceptional overall performances. The change in median overall performance, $\Delta$MOP, for each of the single-channel networks are minor with respect to the $96.0\%$ performance of the three-channel, full-colour \textsc{Ph} networks from Figure \ref{fig:Ph_Ph}: $\Delta$MOP$(r, i)=(-0.4, -0.5)\%$. 

However, there is a potential problem with using the idealized images as a ``representative'' scenario for examining the importance of colour: the possibility of interplay between realism and colour. Between the three realism levels (idealized, \textsc{SemiReal}, \textsc{FullReal}), networks trained using idealized images are least likely to exploit colour information because the low surface brightness morphological features are most easily exploitable. In contrast, low surface brightness morphological features are often hidden in the sky noise or blurred by the PSF in the more realistic \textsc{SemiReal} and \textsc{FullReal} images. Consequently, a colour-sensitive network that is trained using these more realistic images is more likely to use colour to classify galaxies if the correlation between colour and merger-stage is strong \emph{and} morphological information is limited. 

To evaluate the importance of colour in the more realistic image data, we compare the single-channel \textsc{PhFR} test results (see Appendix \ref{app:NC1}) with the three-channel, colour-sensitive \textsc{PhFR} network results (the right panel of Figure \ref{fig:realism}, MOP$=87.1\%$). The \textsc{PhFR} networks trained using individual bands have mild losses in MOP with respect to the three-channel networks: $\Delta$MOP$(r, i)=(-1.1, -2.3)\%$. While these are greater losses than in the idealized case, these results still demonstrate that colour is not an essential ingredient to network success and that the network is primarily targeting morphological features. However, without a training set which includes both red and blue discs, it is uncertain whether it is not simply this information that the network is exploiting to achieve the mildly higher performance in the full-colour data.

Lastly, we compare the timing histograms of both correctly and incorrectly classified galaxies for the single-channel networks (Figure \ref{fig:NC1}) and three-channel networks (Figure \ref{fig:Ph_Ph}). The results are qualitatively similar in every case -- \emph{including the timing preference for isolated galaxies which are misclassified as pairs}. The fact that the timing preference persists in the single-channel networks disqualifies colour as being a driver of preferential misclassification of these isolated galaxies as pairs -- in particular, those corresponding to early, high SFR, and morphologically unstable snapshots from the isolated simulation runs. A network that does not have access to colour cannot exploit its relationship with star-formation. Combined with our visual analysis of correctly and incorrectly classified isolated galaxies in Section \ref{sec:Ph_Ph} and Figure \ref{fig:Iso}, this result demonstrates that the networks are focusing on morphological features. However, it should still be noted that certain morphological properties can be significantly enhanced by star-formation (for example, compactness). In our discussion of the limitations of our suite, we argue that this problem can be solved using cosmological training sets which include greater variety of isolated galaxies and merger properties (gas fractions, initial morphologies, etc.).


\subsubsection{Does the bandpass make a difference?}\label{sec:bandpass}

Table \ref{tab:SemiReal} shows that the typical sky surface brightness uncertainty, $\langle \sigma_{\mathrm{sky,Field}} \rangle$, in the SDSS $r$-band is 1.5 mag/arcsec$^2$ fainter than in the $i$-band. By comparing the performances in each of these bands individually, we determine the sensitivity of network performance to a modest change in imaging depth and the differences in the intrinsic brightnesses of targets in each band. Figure \ref{fig:NC1} shows that the difference in median overall performance in the $i$-band with respect to the $r$-band is only $-0.1\%$ in the idealized \textsc{Ph} images, as expected. The idealized images contain no noise -- so the only change between the $r$-band and the $i$-band are the brightnesses of stellar populations in each bandpass. The difference in performance broadens for networks trained on the more realistic single-channel \textsc{PhFR} images. The single-channel \textsc{PhFR} networks achieve median overall performances of $86.0\%$ ($r$-band) and $84.8\%$ ($i-$band) with a minor difference in MOP of $-1.2\%$ in the $i$-band with respect to the $r$-band. While this is a small change in performance, one must recall that source surface brightness in a bandpass diminishes rapidly with source redshift -- by a factor of $(1+z)^{-5}$. Consequently, a difference of 1.5 mag/arcsec may be a much greater hinderance to network performance for images of galaxies at the higher-$z$ end of a realistic SDSS redshift distribution. 

\section{Discussion}\label{sec:discussion}

\subsection{The importance of realism}

In Section \ref{sec:realism} and \ref{sec:level} and Figures \ref{fig:realism} and \ref{fig:level}, we showed that adding full realism to synthetic training images (including realistic skies, resolution, and crowding by nearby sources) is a necessary condition of strong network performance in identifying and characterizing galaxy interactions in realistic images. Indeed, we have shown that every ingredient of full observational realism is essential and that omitting any ingredient leads to systematic misclassifications in testing. Without exposure to contamination by nearby sources in the training images, networks systematically preferred to assign images to the pair class. In particular, the systematic misclassification of fully realistic images as pairs persists even when the training images have both realistic skies and resolution but lack crowding effects. 

Similarly, the tests in which we train on idealized \textsc{Ph} or \textsc{Sm} images and test on \textsc{PhSR} or \textsc{SmSR} images (orange and blue bars in the two middle panels of Figure \ref{fig:testall}) reveal that realistic skies and resolution are also vital (also see Figure \ref{fig:megamatrix} of Appendix \ref{sec:megamatrix}). The sky and resolution effects bury and wash away the low surface-brightness morphological features that made idealized \textsc{Ph} or \textsc{Sm} networks successful on images of their own types (e.g. tidal tails, bridges and shells). Consequently, networks that are trained using idealized images perform very poorly on test images that contain realistic skies and resolution -- even without crowding effects.\footnote{Note that this simultaneously demonstrates that the realism \emph{should} be survey-specific. However, the results of \cite{2019MNRAS.484...93D} for morphological classifications demonstrate that it may be possible to use a \emph{transfer learning} approach -- in which CNNs optimized for one survey can be adapted to another using a small sample of images from the target survey.} Combined with our results showing the independent biases that arise from excluding crowding effects in training images, the importance of each component of a \textsc{FullReal} image is clear. 

Indeed, in Section \ref{sec:realorrad}, we demonstrated that the level of realism is even more important than whether the synthetic images originate from radiative transfer or from \textsc{StellarMap} images. The bottom right panel of Figure \ref{fig:testall} shows that the only networks with performances that approach those of the \textsc{PhFR} networks on the \textsc{PhFR} test images are the \textsc{SmFR} networks -- which exploit neither colour nor any higher-order features available from radiative transfer. Ultimately, we want to be able to construct training images from the current state-of-the-art cosmological hydrodynamical simulations which will be used to train networks that can then identify and characterize real galaxy interactions by merger-stage. Only these simulations can provide the necessary scope, diversity and accuracy in galaxy properties (e.g. morphologies, masses, merger characteristics, orbital properties, gas fractions, etc.) to form training sets which are sufficiently generalizable to real galaxies and interactions. Consequently, training sets generated from these simulations will necessarily comprise very large numbers of synthetic images. However, in Section \ref{sec:SMvsPh}, we noted that constructing training sets on these scales using radiative transfer corresponds to potentially enormous computational and data management expenses. Therefore, the results that (1) radiative transfer is secondary to realism and (2) one can avoid radiative transfer using \textsc{SmFR} images for a modest compromise in performance are of great importance to the primary science application of the methods we present. 

Lastly, with more diverse data (for example, data that are not so tightly temporally correlated) realism may become an even more important factor in generating large samples of images and training networks which effectively handle crowding effects. For the goals of this paper, it was sufficient to insert each projection into a single field of view and increase the size of our datasets by augmentation. However, one could produce a much larger sample of images and train a network that may be even less sensitive to crowding effects by inserting \emph{each} projection of a synthetic image into, for example, $N=10$ unique fields. This would expose the network to a greater diversity of crowding effects for each target. As long as the training set is sufficiently large and each target gets an equal amount of unique insertions, the network will not overfit to a particular target. This would simultaneously ensure that no overfitting can ever occur due to particular configurations of targets and projected objects in the image FOV.

\subsection{Limitations of the suite}\label{sec:limitations}

As we have previously stated, the \emph{specific} networks we trained in this study would have limited application to real data. This is primarily because our simulations are not cosmological. Despite the fact that the \textsc{FIRE} merger suite covers a broad range of mass ratios and orbital properties, the range is small in comparison to the parameter spaces of galaxy and merger properties encompassed by the observable Universe or by the current state-of-the-art cosmological hydrodynamical simulations (Patton et al. in prep, Blumenthal et al. in prep). Consequently, we know that our training data are not generalizable to real data. Only cosmological simulations can provide the necessary scope to construct training data that are \emph{both} similar and diverse enough to train networks that can be applied to real data.

However, generating networks that can be applied to real data is not a goal of this work. The goals are (1) to provide the methodology with which convolutional neural networks, trained and calibrated using hydrodynamical simulations, can be used to identify mergers and predict merger stage in \emph{realistic} images and (2) to assess the importance of realism in the synthetic training images. The experiments we used to accomplish these goals did not require cosmological simulations. Indeed, our experiments correspond to scenarios in which we know that the training data are fully generalizable to the test data, as desired, because both training and test data are drawn from the same merger suite -- just different parts of it. Ultimately, whether we use training data from a cosmological simulation or from our suite would make no qualitative difference to our results regarding the importance of realism. However, it will be important to assess whether the reduction in intrinsic simulation resolution in the cosmological simulations has an additional effect on performance. We will address this question in a follow-up study. 

Another limitation of the suite used in the current work is that all of the galaxies used therein are relatively gas rich disks, and hence not representative of the full morphological complexity seen in the real Universe. This morphology bias would undoubtedly be an issue if we were to apply our networks to real data, but is not a limitation for the goals of this work. However, observations and numerical simulations alike show that mergers between gas-rich discs induce central star-formation in galaxies during the pair phase and in the merger remnant. Therefore, the relationship between merger-stage and star-formation may be exploited by networks that are sensitive to properties related to central star-formation. While we have demonstrated that eliminating colour sensitivity makes an insignificant difference to network performance in Section \ref{sec:colour}, we do not rule out a morphological connection with high central star-formation rates -- such as with $CAS$ Concentration index \citep{2000AJ....119.2645B,2003ApJS..147....1C} or Gini coefficient \citep{2003ApJ...588..218A,2004AJ....128..163L}. However, the increases in central surface-brightnesses from recent star-formation are associated with the low $M/L$ of young stellar populations formed in the bursts. So while the \textsc{Photometry}-based images and networks are more liable to exploit such connections between recent central star-formation and morphology, the \textsc{StellarMap}-based images will be largely insensitive to the morphological effects of recent star-formation because they are \emph{completely} insensitive to $M/L$ ratio. Figure \ref{fig:testall} showed that the \textsc{SmFR} networks performs nearly as well as the \textsc{PhFR} network. Indeed, a median overall performance of $95.8\%$ is achieved with networks trained and tested on the idealized \textsc{Sm} images compared to the \textsc{96.0\%} achieved by the networks trained and tested on idealized \textsc{Ph} images. Therefore, while a connection between central morphology (as induced by central star-formation) and merger stage may exist in the \textsc{Photometry}-based images, it is not essential to network performance. Additionally, the exploitation of such a connection would be expected to be further suppressed in a more homogeneous galaxy sample (such as from a cosmological simulation) with mergers between galaxies that are red, blue, gas-rich, gas-poor and everything between.

\subsection{Overfitting}

As explained in the previous section, we know that our networks are limited to the set of merger scenarios encompassed by our suite with respect to galaxy and merger properties that would be present in a representative volume of the Universe. However, for evaluating the importance of realism, this limitation is immaterial because all we needed was a reasonably sized training set which includes typical merger features and test sets to which the training data are known to be generalizable. In contrast, a bias that would not be desirable is one that might arise from the construction of our images -- such as camera angles or orientation. For example, in Figure \ref{fig:Iso}, the correctly classified isolated galaxy image in the 3rd row, 2nd column of the left panel is the same galaxy and snapshot as the one that is incorrectly classified as a pair in the 3rd row, 1st column of the right panel. The only difference between these images is a slight change in zoom and rotation. A high sensitivity of predicted class to orientation is a common characteristic of overfitting -- where a network learns to exploit properties of the training data that are not generalizable to test data. 

While CNNs with max-pooling layers are architecturally invariant to translation, they are not rotationally invariant by default and require large and diversified training data to achieve \emph{learned} rotational invariance (see Chapter 9, Figure 9.9 of \citealt{goodfellow2016deep} for an intuitive example). We apply rotational, translational, and zoom augmentation to all of our datasets in an effort to (1) increase our data size and (2) achieve rotationally invariant networks. Given that every image in the augmented training data (including all possible orientations) contributes equally to network optimization, we find it unlikely that our networks are classifying based on orientation. However, another example from Figure \ref{fig:Iso} is the correctly classified inclined disc in the 3rd row, 4th column of the left panel and its incorrectly classified counterpart in the 1st row, 2nd column of the right panel. Both images correspond to the same galaxy and inclination -- only the incorrectly classified one is from a much later snapshot and is rotated. Despite the visual similarity between these targets, the network confidently classifies these images as isolated and pair, respectively. Although Figure \ref{fig:Ph_Ph} shows that such misclassifications are rare, this high sensitivity between isolated and pair classifications, without obvious visible justification, may arise from our class definitions.

\subsection{Class Definitions}\label{sec:disc_classes}

By using hydrodynamical simulations to train networks, we attempt to eliminate as much subjectivity as possible for merger stage classifications. The advantage of this strategy is that, based on a set of simple quantitative definitions for each class, one is always optimizing network performance on the absolute truth. However, the definitions themselves are one remaining source of subjectivity that cannot be avoided in supervised learning. The beginning of the post-merger class requires a definition of coalescence which also defines the end the pair class. The beginning of the pair phase also requires a definition. We defined the pair phase as beginning 100 Myr before first pericentric passage. Was our choice to use this temporal criterion appropriate? What were the consequences? 

Figure \ref{fig:Ph_Ph} and the right panel of Figure \ref{fig:realism} show that pairs that are misclassified as isolated are preferentially \emph{early} pairs. The clear consequence of our definition is that galaxies in the early pair phase are indistinguishable from isolated galaxies because no galaxies in these early pairs have experienced visible disturbances resulting from gravitational interaction with their companions. Subsequently, this definition is also a likely culprit for the seemingly spurious misclassifications of a few isolated galaxies shown in Figure \ref{fig:Iso} that were discussed in the last section. However, for the purposes of this paper, our definition happened to be beneficial (see Section \ref{sec:Ph_Ph}). The fact that the networks had difficulty distinguishing early pairs (by our definition of the pair phase) from isolated galaxies was evidence that the networks were behaving intuitively. Meanwhile, since the majority of images from the pair class do not resemble isolated galaxies (all those except for the early pairs), the networks still accurately classified most pairs in the test images. 

Ultimately, we propose that reduced continuity between the isolated and pair images through an alternative definition of the pair phase would lead to better network performance and fewer misclassifications in these classes (for example, starting the pair phase at first pericentre). However, testing the sensitivity of performance to alternative definitions for each class is beyond the scope of this work. There is a large parameter space to be explored. The time or spatial separation at which a galaxy's properties start to be affected by an interaction and persist after coalescence are sensitive to the masses, morphologies, mass ratio and orbital properties at hand (e.g. \citealt{2008MNRAS.391.1137L,2010MNRAS.404..575L,2010MNRAS.404..590L,2014A&A...566A..97J,2019ApJ...872...76N}). Nonetheless, we highlight that a few key advantages of calibrating networks using simulations is that, for a given set of class definitions, one can (1) train networks which make completely reproducible predictions and (2) evaluate the biases associated with these definitions. So, while \emph{our} class definitions resulted in some confusion between early pairs and the isolated class, these definitions can be easily changed and optimized to improve performance. 
 
\section{Summary}\label{sec:summary}

Convolutional neural networks are becoming a popular tool for identifying galaxy mergers in large surveys. In this paper, we use galaxy merger simulations to train CNNs which identify mergers and predict merger stage. We assess the importance of producing realistic images from simulations to the performance of CNNs, in order to guide future applications of this method. 

We train and calibrate a set of convolutional neural networks using synthetic images generated from a suite of hydrodynamical binary merger simulations \citep{2019MNRAS.485.1320M} run with the \textsc{FIRE-2} physical model \citep{2018MNRAS.480..800H}. Training networks on simulations offers the significant benefit of foreknowledge of interaction stage and, therefore, optimization targets that are not biased by factors such as image quality or personal subjectivity. We examine the importance of adding realistic ingredients to the synthetic images. To do so, networks are trained using two types of galaxy images, stellar maps and dust-inclusive radiatively transferred images, each with three levels of observational realism: (1) no observational effects (idealized images), (2) realistic sky and point spread function (semi-realistic images) and (3) insertion into a real sky image (fully realistic images) (see Section \ref{sec:types} and summary in Table \ref{tab:types}). Each image dataset covers the same set projections and simulation snapshots and is divided into isolated, pair, and post-merger classes. In our main handshake experiment (see Section \ref{sec:handshake} and Figure \ref{fig:handshake}), we test each network on data of every other type. Each network is also tested on data of the same type upon which it was trained but that the network never sees during training. The \textsc{PhFR} data -- in which the synthetic images are injected into real survey fields -- are the most realistic representation of real observations. Therefore, the \textsc{PhFR} test data are used to evaluate how well networks trained on images of a particular type would handle real data (see Section \ref{sec:limitations} for an important discussion on the limitations of \emph{this suite} for applications to real data). The results of our main handshake experiment are:
\begin{enumerate}

\item $[$Section \ref{sec:Ph_Ph}$]$ \textbf{Networks trained on idealized images (\textsc{Sm} and \textsc{Ph}) classify images of the same type with $96.0\%$ accuracy} (Figure \ref{fig:Ph_Ph} and the upper left panel of Figure \ref{fig:SM_Ph}). Misclassifications behave predictably. Early pairs are difficult to distinguish from isolated galaxies. Recent post-mergers are difficult to distinguish from pairs nearing coalescence. Isolated galaxies and post-mergers are \emph{never} confused for one-another. \\

\item $[$Section \ref{sec:SMvsPh}$]$ \textbf{\textsc{Sm} images can be used in place of more computationally expensive (but more realistic) images produced with radiative transfer at a modest cost in performance.} Networks trained on idealized \textsc{Sm} images classify idealized \textsc{Ph} images with $90.2\%$ accuracy (upper right panel of Figure \ref{fig:SM_Ph}). \\ 

\item $[$Section \ref{sec:realism}$]$ \textbf{\textsc{Ph} and \textsc{PhSR} networks -- of which neither are exposed to training images which include contamination by nearby sources -- systematically classify \textsc{PhFR} images as belonging to the pair class.} Networks trained on idealized \textsc{Ph} or semi-realistic \textsc{PhSR} images (realistic skies and resolution) both perform very poorly ($48.9\%$ and $56.2\%$ accuracies, respectively) on the \textsc{PhFR} images (left and centre panels of Figure \ref{fig:realism}). \\

\item $[$Section \ref{sec:level}$]$ \textbf{As long as networks are exposed to all ingredients of realism in training (skies, resolution and crowding) they can learn to efficiently handle these effects in test images.} While \textsc{Ph} and \textsc{PhSR} networks fail to handle realistic images, networks trained on \textsc{PhFR} images classify \textsc{PhFR} test images with $87.1\%$ accuracy (right panel of Figure \ref{fig:realism}). Additionally, (a) there is no clear systematic preference toward classifying images as pairs and (b) \textsc{PhFR} networks are even more accurate on \textsc{PhSR} test images ($88.4\%$) than \textsc{PhFR} test images (see Figure \ref{fig:level}). \\

\item $[$Section \ref{sec:realorrad}$]$ \textbf{Realism is more important than whether the images originate from radiative transfer or from maps of stellar mass.} Networks trained on \textsc{SmFR} images classify \textsc{PhFR} test images with $79.6\%$ accuracy (lower right panel Figure \ref{fig:testall}). Indeed, these are the only networks other than the \textsc{PhFR} networks that achieve reasonable performance on the \textsc{PhFR} test images.

\end{enumerate}

We perform a secondary handshake experiment aimed at characterizing the roles of colour and depth to network performance (see Section \ref{sec:NC1}). Single-channel networks are trained on the $r$ and $i$ band images, individually, taken from the \textsc{Ph}, \textsc{PhSR} and \textsc{PhFR} datasets. These tests eliminate the possibility for networks to exploit colour information and allows us to compare results for networks trained on images in bands of varying photometric depths. The main results of these tests are:

\begin{enumerate}

\item $[$Section \ref{sec:colour}$]$ \textbf{Networks trained without colour incur very mild penalties to performance with respect to colour-sensitive networks}. The performances of the single-channel $r$- and $i$-band \textsc{Ph} (\textsc{PhFR}) networks are $95.6\%$ ($86.0\%$) and $95.5\%$ ($84.8\%$), respectively, compared to $96.0\%$ ($87.1\%$) with the full-colour networks (see Figure \ref{fig:NC1}). These results demonstrate that, while the colour-sensitive networks \emph{can} exploit colour information, colour is not a \emph{necessary} ingredient for high network performance. \\

\item $[$Section \ref{sec:bandpass}$]$ \textbf{The difference in average photometric depth between the $r$- and $i$-bands ($\sim 1.5$ mag/arcsec$^2$) yields a small difference in the performances of networks trained on each band individually} (see Figure \ref{fig:NC1}). However, in this study, we do not match the redshift distribution of SDSS galaxies and instead insert galaxies at the median redshift of galaxies in the DR14 MaNGA galaxy sample. Therefore, these differences might be expected to be larger for training and test data which include galaxies that are more distant or have lower intrinsic brightnesses.
\end{enumerate}

The pertinent applications of this work are: (1) to train networks using realistic synthetic images from cosmological simulations and (2) to use a model trained on cosmological simulations to identify and characterize interactions in the real Universe. The most important feature of cosmological simulations in this respect is that mergers \emph{and} isolated galaxies that are selected from a statistically representative simulation will cover a larger range of morphologies, masses, gas fractions, etc. This diversity will be a necessary component of a training set that can be expected to perform well on real test data.

\section*{Acknowledgements}

CB acknowledges the support of a National Sciences and Engineering Research Council of Canada (NSERC) Graduate Scholarship. MHH and HT contributed equally to this research. MHH acknowledges the receipt of a Vanier Canada Graduate Scholarship. SLE and LS acknowledge the support under the Canadian Discover Grants Program. The data used in this paper were, in part, generated and hosted using facilities supported by the Scientific Computing Core at the Centre for Computational Astrophysics, a division of the Simons Foundation. The computations in this research were enabled in part by support provided by Compute Canada (\href{www.computecanada.ca}{www.computecanada.ca}). The numerical simulations in this paper were run on the Odyssey cluster supported by the FAS Division of Science, Research Computing Group at Harvard University. Support for JM is provided by the NSF (AST Award Number 1516374), and by the Harvard Institute for Theory and Computation, through their Visiting Scholars Program. 

Funding for the Sloan Digital Sky Survey IV has been provided by the Alfred P. Sloan Foundation, the U.S. Department of Energy Office of Science, and the Participating Institutions. SDSS-IV acknowledges
support and resources from the Center for High-Performance Computing at
the University of Utah. The SDSS web site is \href{www.sdss.org}{www.sdss.org}.

SDSS-IV is managed by the Astrophysical Research Consortium for the 
Participating Institutions of the SDSS Collaboration including the 
Brazilian Participation Group, the Carnegie Institution for Science, 
Carnegie Mellon University, the Chilean Participation Group, the French Participation Group, Harvard-Smithsonian Center for Astrophysics, 
Instituto de Astrof\'isica de Canarias, The Johns Hopkins University, Kavli Institute for the Physics and Mathematics of the Universe (IPMU) / 
University of Tokyo, the Korean Participation Group, Lawrence Berkeley National Laboratory, 
Leibniz Institut f\"ur Astrophysik Potsdam (AIP),  
Max-Planck-Institut f\"ur Astronomie (MPIA Heidelberg), 
Max-Planck-Institut f\"ur Astrophysik (MPA Garching), 
Max-Planck-Institut f\"ur Extraterrestrische Physik (MPE), 
National Astronomical Observatories of China, New Mexico State University, 
New York University, University of Notre Dame, 
Observat\'ario Nacional / MCTI, The Ohio State University, 
Pennsylvania State University, Shanghai Astronomical Observatory, 
United Kingdom Participation Group,
Universidad Nacional Aut\'onoma de M\'exico, University of Arizona, 
University of Colorado Boulder, University of Oxford, University of Portsmouth, 
University of Utah, University of Virginia, University of Washington, University of Wisconsin, 
Vanderbilt University, and Yale University.




\bibliographystyle{mnras}
\bibliography{References/References.bib}



\appendix

\section{Single-band photometry results}\label{app:NC1}

\begin{figure}
	\includegraphics[width=\linewidth]{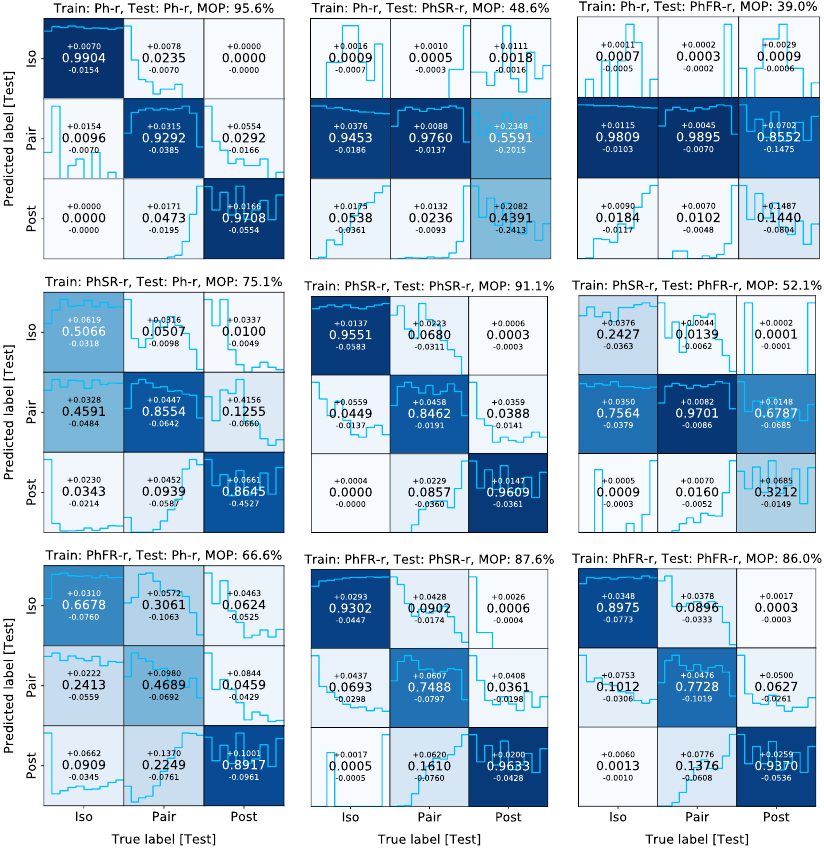}

    \caption{Confusion matrices for the single-channel handshake experiment in which the $r$-band images from \textsc{Ph}, \textsc{PhSR} and \textsc{PhFR} datasets are used to train colour-insensitive networks which are then applied to images of each type. For reference, Figure \ref{fig:Ph_Ph} describes the information displayed by each individual confusion matrix in detail. }
    \label{fig:singlechannel_r}
\end{figure}

\begin{figure}
	\includegraphics[width=\linewidth]{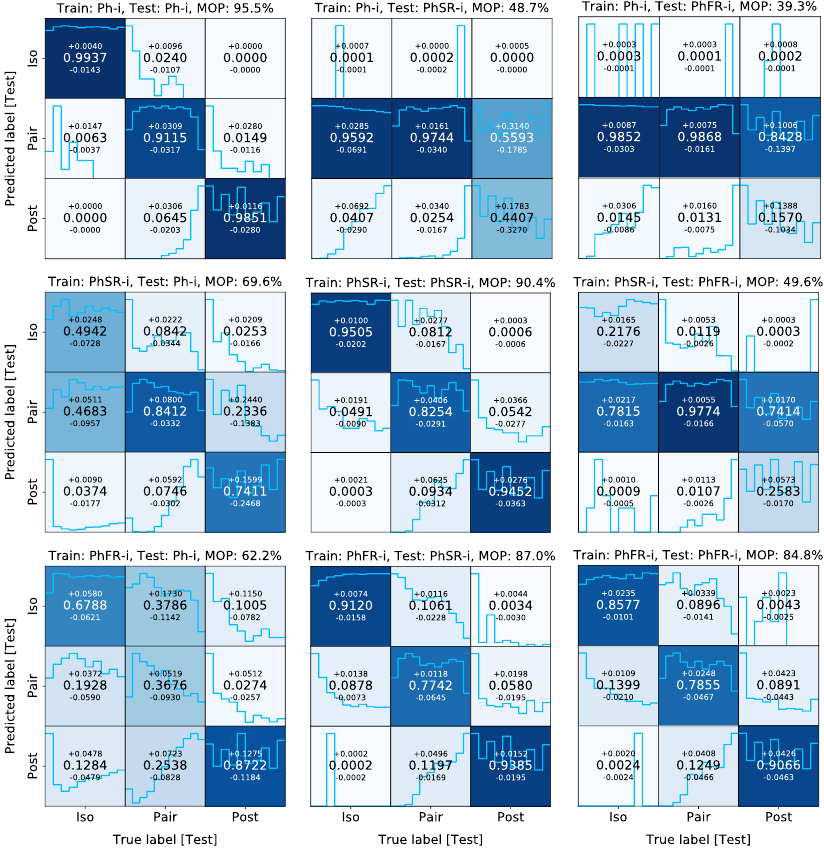}
    \caption{Same as Figure \ref{fig:singlechannel_i} but for networks trained and tested on $i$-band images from the \textsc{Ph}, \textsc{PhSR} and \textsc{PhFR} datasets.}
    \label{fig:singlechannel_i}
\end{figure}

Figures \ref{fig:singlechannel_r} and \ref{fig:singlechannel_i} show the results for the single-channel handshake experiment in which the $r$-band and $i$-band images from \textsc{Ph}, \textsc{PhSR} and \textsc{PhFR} datasets are used to train single-channel, colour-insensitive networks which are then applied to images of each type.

\section{Correlations between galaxy images}\label{sec:corr}

In this appendix, we investigate the possibility that our snapshot sampling cadence for each interaction (e.g. see Figure \ref{fig:selection}) was too fine -- resulting in galaxy images that could be strongly correlated. For our datasets, strong correlations between images in neighbouring snapshots could result in overfitting to the training data and, consequently, loss of generalizability to test data from other datasets (e.g. train: \textsc{Ph}, test: \textsc{PhFR}). Meanwhile, strong correlations between images in tests where a network is applied to test data from the same dataset (e.g. train: \textsc{Ph}, test: \textsc{Ph}) may lead to correlated training/test data and result in erroneously high test accuracies. Since our main investigation is focused on the sensitivity of network performance to realism, it is crucial to characterize (and preferably rule out) the sensitivity of the network performance to our snapshot selection cadence. 

\begin{figure*}

	\includegraphics[width=\linewidth]{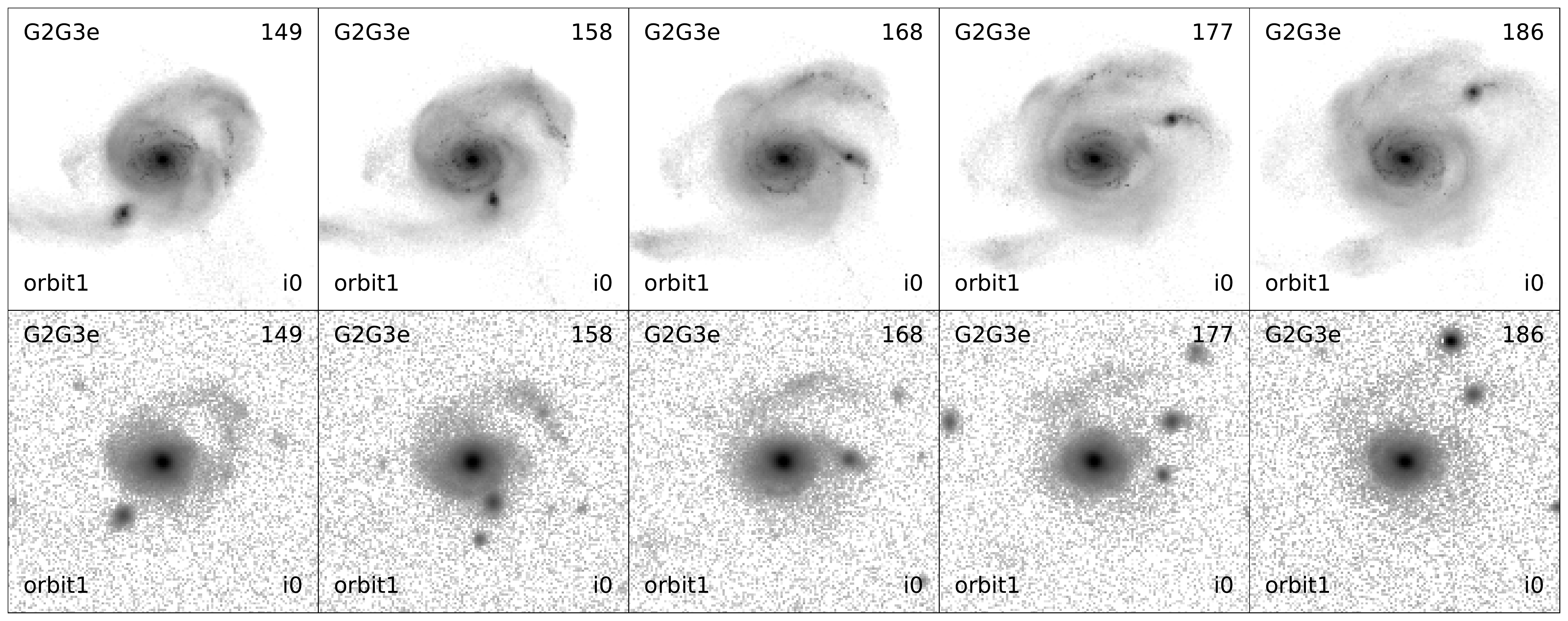}
    \caption{\textsc{Ph} (top row) and corresponding \textsc{PhFR} (bottom row) images for five neighbouring snapshots at fixed camera angle from our sampling of the G2G3e orbit 1 merger near first apocenter ($t\approx0.75$ Gyr in Figures \ref{fig:selection} and \ref{fig:corr_seq}). If there were strong correlations between images from neighbouring snapshots, they would be most likely to occur here, at first apocenter in the merger sequence -- since the rate of morphological evolution should be lowest relative to the more rapid changes at first pericenter and beyond second pericenter. Visually, inner and outer structures of the galaxy both evolve appreciably in this sequence of images. These visual differences are apparent in both the \textsc{Ph} and \textsc{PhFR} images. Figure \ref{fig:corr_seq} shows the results of a more quantitative and robust experiment which demonstrates that images such as these are not so strongly correlated that they are influencing our main results.}
    \label{fig:corr_ex}
\end{figure*}

Figure \ref{fig:corr_ex} shows \textsc{Ph} (upper panels) and corresponding \textsc{PhFR} images (lower panels) for five neighbouring snapshots from our sampling of the fiducial G2G3e orbit 1 merger at a fixed camera angle. The snapshots are selected from near first apocenter ($t\approx0.75$ Gyr in Figure \ref{fig:selection}) where the rate of morphological evolution would be expected to be at a minimum during the pair phase and, therefore, are most likely to be visually correlated. It should be noted that the smaller object in the images is not the companion (which is outside the field of view) but is a tidal dwarf that is produced in this particular interaction. The \textsc{Ph} images in the upper panels of Figure \ref{fig:corr_ex} show that there is still visual evolution in galaxy structure (both inner and outer) during this relatively calm part of the pair phase. The \textsc{PhFR} images in the lower panels are even less visibly correlated due to the varying sky levels, resolution, and contamination by additional sources. Correlations between the images should therefore stand to be most problematic in the idealized images which do not incorporate any additional realism. However, since these visual assessments are subjective, we also devised a test which quantitatively investigates the possibility of correlations between images from neighbouring snapshots.

If images from neighbouring snapshots are strongly correlated, then randomly sampling the test data from a dataset (e.g. splitting \textsc{Sm} into training/validation/test data) will result in training, validation, and test data that are correspondingly correlated. There are two experiments which could be performed to investigate whether such correlations may be influencing our results: (1) reduce the sampling cadence (e.g. discard every second snapshot from our initial selection) and (2) reserve all images from a full merger simulation for testing and train/validate on the remaining images. The problem with (1) is that this approach would reduce our data volume by an integer factor -- which could lead to overfitting from sparseness in the dataset and make it impossible to assign any reduction in accuracy to either this sparseness or image correlations. Test (2) does not suffer from this problem because, for a given merger, it only reduces the data volume by relatively small factor of $1/23$. Poor classification performance on the images from the merger that were reserved for testing (particularly for phases where the rate of morphological evolution is expected to be small) would confirm that the snapshot selection is too fine and yields correlations between images that would be affecting our results. The second test would simultaneously show whether a network is generalizable to images from a merger that it has never seen before (albeit, within the limitations outlined in Section \ref{sec:limitations}). For these reasons, we carry out the second test. 

\begin{figure*}

	\includegraphics[width=0.69\linewidth]{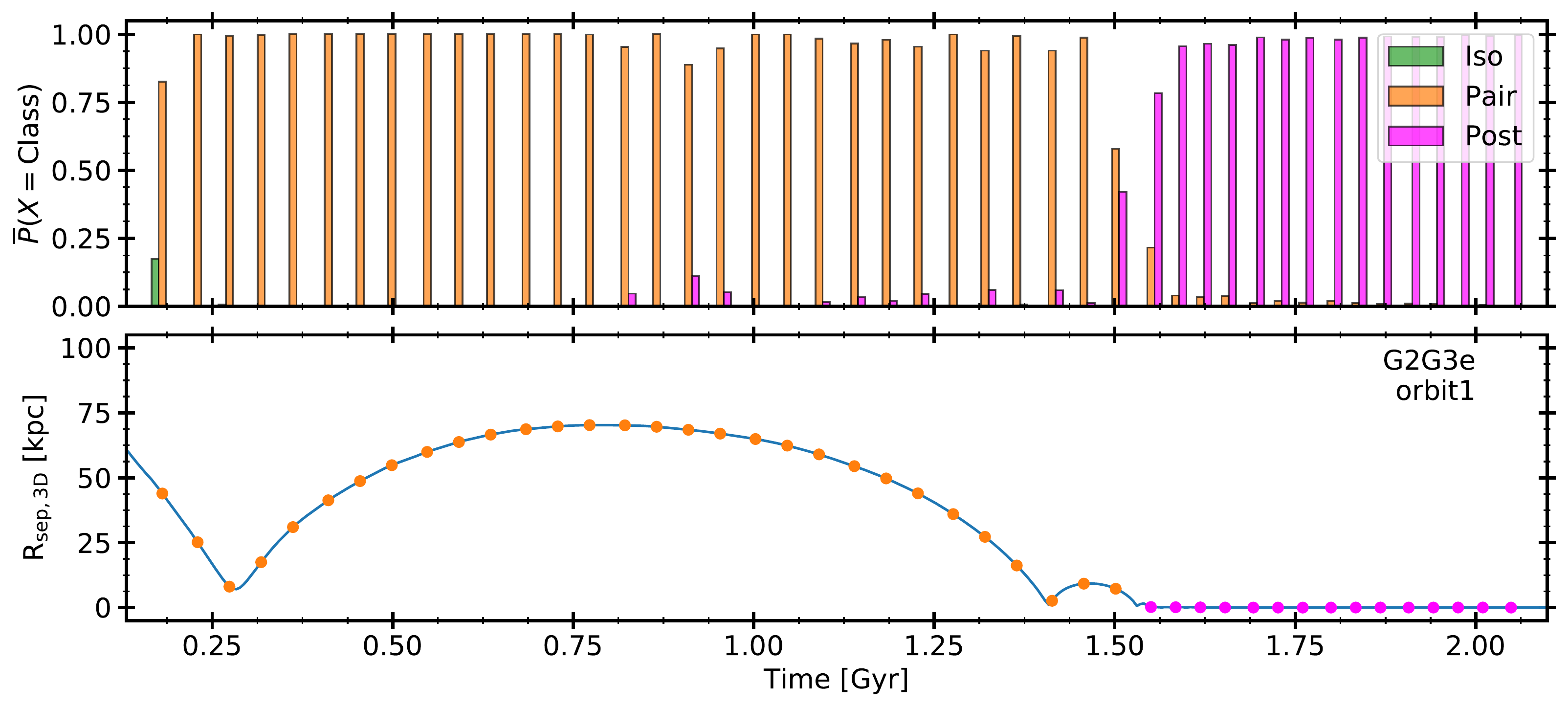}
	\includegraphics[width=0.30\linewidth]{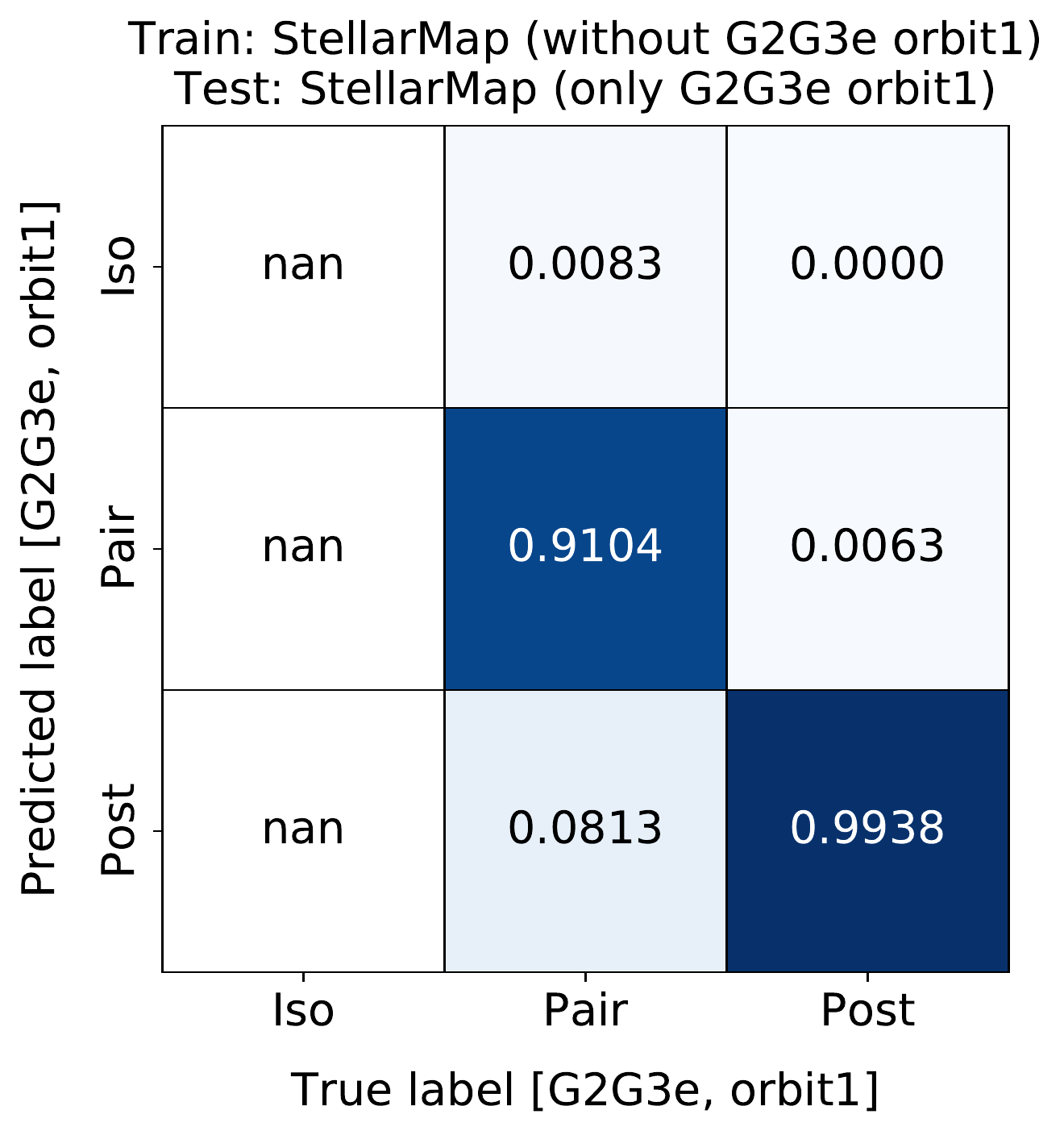}
    \caption{Classification scores along fiducial G2G3e orbit 1 interaction sequence (left panels) and confusion matrix (right panel) for the experiment described in Appendix \ref{sec:corr}. The experiment is designed to show whether the sampling cadence for each merger yields images that are too strongly correlated -- resulting in test images that are too similar to the training/validation images. Applying a network to images from an \emph{entire merger} that the network never saw during training would reveal whether such correlations are affecting our results. The images from the fiducial G2G3e orbit 1 merger were removed from the \textsc{Sm} dataset and reserved for testing. A \textsc{Sm} network was then training using the remaining data \textsc{Sm} and applied to the G2G3e orbit 1 test images. The right panel shows the confusion matrix for this test. These results are consistent with the results shown in the upper left panel of Figure \ref{fig:SM_Ph} (train: \textsc{Sm}, test: \textsc{Sm}) -- demonstrating that possible correlations between images from neighbouring snapshots are not affecting our results in the main experiments. The left panels show the classification scores as coloured bars for each snapshot in the G2G3e orbit 1 test sequence, averaged over all camera angles at each snapshot, $\overline{P}(X=\mathrm{Class})$. The results of this test show that: (1) that the snapshot sampling cadence is not affecting our networks' performances even in phases where the rate of morphological evolution is expected to be minimal (e.g. near apocenter) and (2) there is a continuous transition between class scores at the temporal class boundaries (e.g. between Pair and Post-merger at $t\approx1.54$ Gyr).}
    \label{fig:corr_seq}
\end{figure*}

Figure \ref{fig:corr_seq} shows the results of our investigation into possible correlations between images from neighbouring snapshots. The investigation was performed using the \textsc{Sm} dataset. We removed all images from the fiducial G2G3e orbit 1 merger (Figure \ref{fig:selection}) from the \textsc{Sm} dataset and reserved them as a test set. We also removed images from the G2G3 merger in the mass ratio suite (which all use ``e'' orbit 1 initial conditions). This merger is not identical to the fiducial simulation (due to the chaotic nature of galaxy mergers) but has the same initial conditions and is consequently removed and discarded to eliminate any possibility that the training/validation data include images which may be correlated with the test images from the fiducial G2G3e orbit 1 merger simulation. We trained a network on the remaining data with a (70,30)\% training/validation split. Only the images for the fiducial merger were used as test data. The right panel of Figure \ref{fig:corr_seq} shows the confusion matrix for this test. This test set contains no images with the Iso class because those images are drawn from separate isolated simulation runs as outlined in Section \ref{sec:selection}. Consequently, all elements in the first column are NaNs. The performance on the G2G3e orbit 1 test images, after removing all images from the training/validation data which might be correlated to these test images, is consistent with the results shown in the upper left panel of Figure \ref{fig:SM_Ph}. In other words, the network runs equally well on a merger for which it has seen seen zero images (this test), as mergers for which $\sim70\%$ of their images are included in the training data (upper left panel of Figure \ref{fig:SM_Ph}). The results of this test demonstrate that possible correlations between images in neighbouring snapshots do not have a significant role in the successes of our networks.

The upper left panel of Figure \ref{fig:corr_seq} shows the normalized classification scores as coloured bars for each snapshot in the G2G3e orbit 1 test images, averaged over all camera angles at each snapshot, $\overline{P}(X=\mathrm{Class})$. The lower left panel shows the radial separation sequence for this merger with each selected snapshot and its true class indicated with coloured circles as in Figure \ref{fig:selection}. This classification sequence plot demonstrates the two main results of this appendix: (1) correlations between images from neighbouring snapshots is not a significant contributor to the accuracies reported in our main analyses (even near first apocenter, where such correlations would be expected to be strongest) and (2) the changes to classification scores as one approaches a temporal class boundary are continuous and not choppy or sporadic (e.g. between the Pair and Post-merger phase at $t\approx1.54$ Gyr). 

\section{Main handshake results}\label{sec:megamatrix}

\begin{figure*}

	\includegraphics[width=\linewidth]{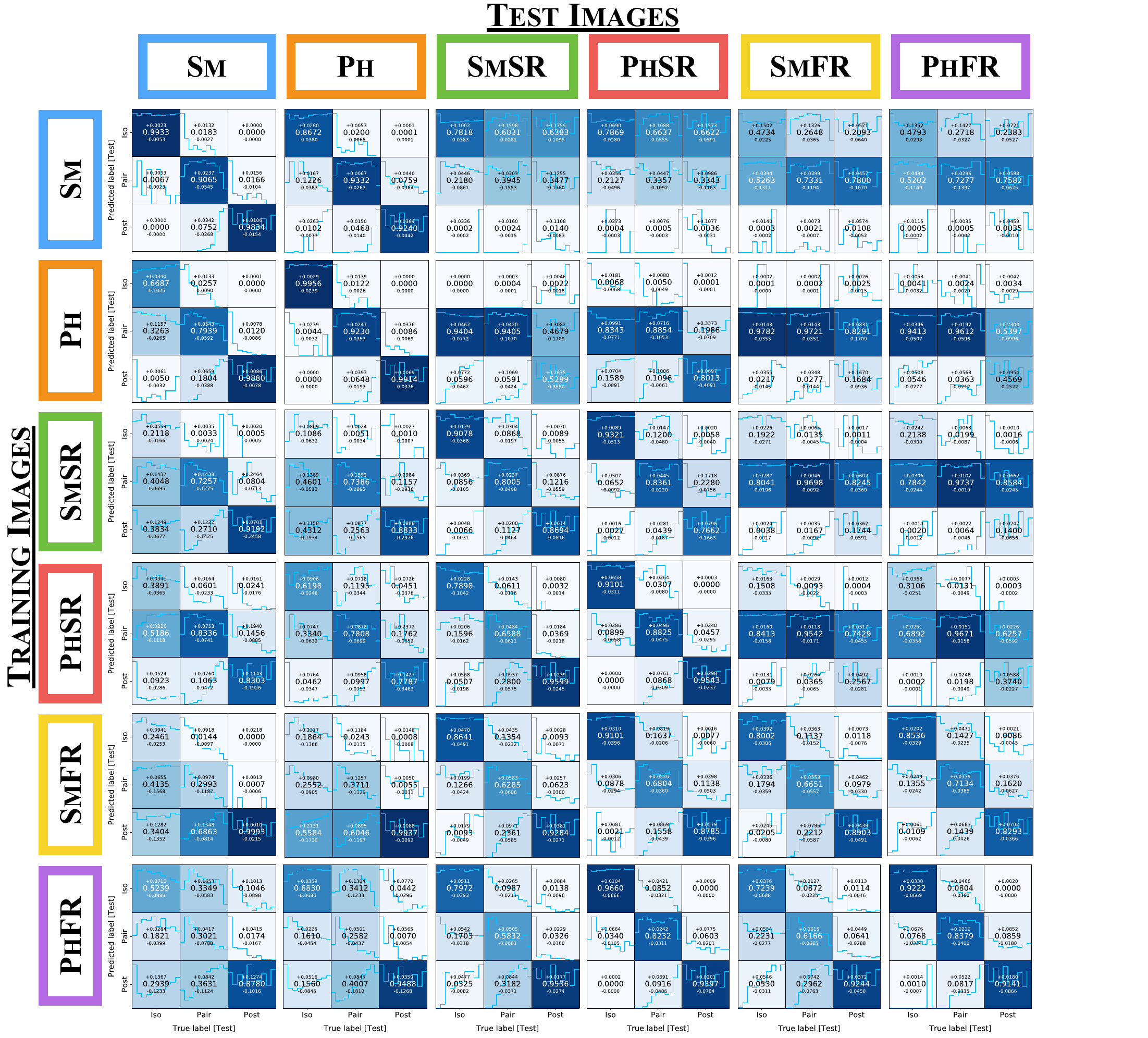}
    \caption{Confusion matrices corresponding to every test carried out in Section \ref{sec:handshake}. For reference, Figure \ref{fig:Ph_Ph} describes the information displayed by each individual confusion matrix in detail. The median overall performance of each result is shown in Figure \ref{fig:testall}.}
    \label{fig:megamatrix}
\end{figure*}

Figure \ref{fig:megamatrix} shows the combined confusion matrices for every test in the main handshake experiment. Each row corresponds to a type of training data. Each column corresponds to a type of test data. As in Figure \ref{fig:Ph_Ph}, each matrix shown combines 10 individual tests performed with different random allocations of training, validation and test images.

\bsp	
\label{lastpage}
\end{document}